\documentclass[prl,twocolumn,amsmath,amssymb,superscriptaddress]{revtex4-2}

\usepackage{physics}
\usepackage{color}
\usepackage{multirow}
\usepackage{gensymb}
\usepackage{graphicx}
\usepackage{multirow}
\usepackage{dcolumn}
\usepackage{bm}
\usepackage[colorlinks=true,linkcolor=blue,citecolor=blue,urlcolor=blue]{hyperref}
\usepackage{float}

\begin{document}

\title{Flexible Readout and Unconditional Reset for Superconducting Multi-Qubit Processors with Tunable Purcell Filters}

\author{Yong-Xi Xiao}
    \thanks{These authors contributed equally to this work.}
\author{Da'er Feng}
    \thanks{These authors contributed equally to this work.}
\author{Xu-Yang Gu}
    \affiliation{Beijing National Laboratory for Condensed Matter Physics, Institute of Physics, Chinese Academy of Sciences, Beijing 100190, China}
    \affiliation{School of Physics, University of Chinese Academy of Sciences Beijing 100049, China}
    
\author{Gui-Han Liang}
    \affiliation{Beijing National Laboratory for Condensed Matter Physics, Institute of Physics, Chinese Academy of Sciences, Beijing 100190, China}
    \affiliation{School of Physics, University of Chinese Academy of Sciences Beijing 100049, China}
\author{Ming-Chuan Wang}
    \affiliation{Beijing National Laboratory for Condensed Matter Physics, Institute of Physics, Chinese Academy of Sciences, Beijing 100190, China}
    \affiliation{School of Physics, University of Chinese Academy of Sciences Beijing 100049, China}
\author{Zheng-Yu Peng}
    \affiliation{Beijing National Laboratory for Condensed Matter Physics, Institute of Physics, Chinese Academy of Sciences, Beijing 100190, China}
    \affiliation{School of Physics, University of Chinese Academy of Sciences Beijing 100049, China}
\author{Bing-Jie Chen}
    \affiliation{Beijing National Laboratory for Condensed Matter Physics, Institute of Physics, Chinese Academy of Sciences, Beijing 100190, China}
    \affiliation{School of Physics, University of Chinese Academy of Sciences Beijing 100049, China}
\author{Yu Yan}
    \affiliation{Beijing National Laboratory for Condensed Matter Physics, Institute of Physics, Chinese Academy of Sciences, Beijing 100190, China}
    \affiliation{School of Physics, University of Chinese Academy of Sciences Beijing 100049, China}
\author{Zheng-Yang Mei}
    \affiliation{Beijing National Laboratory for Condensed Matter Physics, Institute of Physics, Chinese Academy of Sciences, Beijing 100190, China}
    \affiliation{School of Physics, University of Chinese Academy of Sciences Beijing 100049, China}
\author{Si-Lu Zhao}
    \affiliation{Beijing National Laboratory for Condensed Matter Physics, Institute of Physics, Chinese Academy of Sciences, Beijing 100190, China}
    \affiliation{School of Physics, University of Chinese Academy of Sciences Beijing 100049, China}
\author{Yi-Zhou Bu}
    \affiliation{Beijing National Laboratory for Condensed Matter Physics, Institute of Physics, Chinese Academy of Sciences, Beijing 100190, China}
    \affiliation{School of Physics, University of Chinese Academy of Sciences Beijing 100049, China}
\author{Cheng-Lin Deng}
    \affiliation{Beijing National Laboratory for Condensed Matter Physics, Institute of Physics, Chinese Academy of Sciences, Beijing 100190, China}
    \affiliation{School of Physics, University of Chinese Academy of Sciences Beijing 100049, China}
    
\author{Kai Yang}
\affiliation{Beijing National Laboratory for Condensed Matter Physics, Institute of Physics, Chinese Academy of Sciences, Beijing 100190, China}
\affiliation{School of Physics, University of Chinese Academy of Sciences Beijing 100049, China}    
 
\author{Ye Tian}
\affiliation{Beijing National Laboratory for Condensed Matter Physics, Institute of Physics, Chinese Academy of Sciences, Beijing 100190, China}

\author{Xiaohui Song}
    \affiliation{Beijing National Laboratory for Condensed Matter Physics, Institute of Physics, Chinese Academy of Sciences, Beijing 100190, China}
    \affiliation{School of Physics, University of Chinese Academy of Sciences Beijing 100049, China}
    
\author{Dongning Zheng}
    \affiliation{Beijing National Laboratory for Condensed Matter Physics, Institute of Physics, Chinese Academy of Sciences, Beijing 100190, China}
    \affiliation{School of Physics, University of Chinese Academy of Sciences Beijing 100049, China}
    \affiliation{Beijing Academy of Quantum Information Sciences, 100193 Beijing, China}
    \affiliation{Hefei National Laboratory, 230088 Hefei, China}
\author{Yu-Xiang Zhang}
\affiliation{Beijing National Laboratory for Condensed Matter Physics, Institute of Physics, Chinese Academy of Sciences, Beijing 100190, China}
\affiliation{School of Physics, University of Chinese Academy of Sciences Beijing 100049, China}

\author{Yun-Hao Shi}
\affiliation{Beijing National Laboratory for Condensed Matter Physics, Institute of Physics, Chinese Academy of Sciences, Beijing 100190, China}

\author{Zhongcheng Xiang}
    \thanks{zcxiang@iphy.ac.cn}
    \affiliation{Beijing National Laboratory for Condensed Matter Physics, Institute of Physics, Chinese Academy of Sciences, Beijing 100190, China}
    \affiliation{School of Physics, University of Chinese Academy of Sciences Beijing 100049, China}
    \affiliation{Beijing Academy of Quantum Information Sciences, 100193 Beijing, China}
    \affiliation{Hefei National Laboratory, 230088 Hefei, China}
    
\author{Kai Xu}
    \thanks{kaixu@iphy.ac.cn}
    \affiliation{Beijing National Laboratory for Condensed Matter Physics, Institute of Physics, Chinese Academy of Sciences, Beijing 100190, China}
    \affiliation{School of Physics, University of Chinese Academy of Sciences Beijing 100049, China}
    \affiliation{Beijing Academy of Quantum Information Sciences, 100193 Beijing, China}
    \affiliation{Hefei National Laboratory, 230088 Hefei, China}
    \affiliation{Songshan Lake Materials Laboratory, 523808 Dongguan, Guangdong, China}

\author{Heng Fan}
    \thanks{hfan@iphy.ac.cn}
    \affiliation{Beijing National Laboratory for Condensed Matter Physics, Institute of Physics, Chinese Academy of Sciences, Beijing 100190, China}
    \affiliation{School of Physics, University of Chinese Academy of Sciences Beijing 100049, China}
    \affiliation{Beijing Academy of Quantum Information Sciences, 100193 Beijing, China}
    \affiliation{Hefei National Laboratory, 230088 Hefei, China}
    \affiliation{Songshan Lake Materials Laboratory, 523808 Dongguan, Guangdong, China}

\begin{abstract}

Achieving high-fidelity qubit readout and reset while preserving qubit coherence is essential for quantum error correction and other advanced quantum algorithms. Here, we design and experimentally demonstrate a scalable architecture employing frequency-tunable nonlinear Purcell filters, enabling flexible readout and fast unconditional reset of multiple superconducting qubits. Our readout protocol dynamically adjusts the effective linewidth of the readout resonator through a tunable Purcell filter, optimizing the signal-to-noise ratio during measurement while suppressing photon noise during idle periods. We achieve a readout fidelity of $99.3\%$ without any quantum-limited amplifier, even with a small dispersive shift. Moreover, by leveraging a reset channel formed via the adjacent coupling between the filter and the coupler, we realize unconditional qubit reset of both leakage-induced $\ket{2}$ and $\ket{1}$ states within 200 ns and reset of the $\ket{1}$ state alone within 75 ns, with error rates $\leq1\%$. The filter also mitigates both photon-induced dephasing and the Purcell effect, thereby preserving qubit coherence. This scalable Purcell filter architecture shows exceptional performance in qubit readout, reset, and protection, marking it as a promising hardware component for advancing fault-tolerant quantum computing systems.

\end{abstract}

\maketitle

Superconducting qubits have emerged as a leading platform for scalable fault-tolerant quantum computation due to breakthroughs in coherence times, integration, and control fidelity~\cite{oliver2020,wang2022t1,zhang2024,liang2023}. Recent advances have enabled large-scale quantum information tasks, including quantum simulation and error correction~\cite{zhao2022,krinner2022,deng2024,wang2025,google2025}. The growing complexity of these applications demands simultaneous improvements in qubit readout~\cite{sank2016,blais2021,shillito2022,yufeng2024} and reset~\cite{egger2018,nakamura2023reset,maurya2024,yanfei2025}.

Balancing high readout fidelity with qubit protection remains a central challenge in dispersive readout of superconducting qubits~\cite{wallraff2004,koch2007,jeffrey2014,walter2017,heinsoo2018}. While fixed-frequency linear Purcell filters improve signal-to-noise ratio (SNR) and suppress the Purcell effect~\cite{reed2010,jeffrey2014,sank2016,walter2017,heinsoo2018,khezri2023,peter2024}, they cannot fully mitigate photon-noise-induced dephasing. Recent demonstrations of nonlinear Purcell filters on single-qubit devices~\cite{sunada2024} have enabled photon-noise-tolerant readout but rely on large dispersive shifts and remain unverified for scalability. Meanwhile, prior multi-qubit implementations of high-fidelity readout typically depend on quantum-limited amplifiers such as Josephson parametric amplifiers (JPAs) or traveling-wave parametric amplifiers (TWPAs) to achieve sub-1\% readout error~\cite{swiadek2024,bengtsson2024}.

In this Letter, we present a scalable architecture of frequency-tunable nonlinear Purcell filters integrated into a superconducting multi-qubit processor. By dynamically adjusting the effective linewidth of readout resonators through filter tuning, we optimize SNR for qubit state discrimination while suppressing photon noise during idle periods. Critically, this approach achieves 99.3\% readout fidelity without requiring quantum-limitied amplifier, surpassing prior multi-qubit benchmarks under comparable conditions. Our design operates with a small dispersive shift, enabling compatibility with compact, low-noise architectures.

Additionally, qubit reset capable of mitigating leakage errors is crucial for quantum error correction~\cite{wang2024zu,wallraff2025}. Thus, we experimentally implement a coupler-assisted qubit reset protocol that utilizes the filter as a fast dissipation channel. This also demonstrates the functional versatility of our tunable Purcell filter. Our method achieves unconditional reset of leakage-induced $\ket{2}$ and $\ket{1}$ within $200$~ns while reset of $\ket{1}$ alone completes in just 75~ns, with error rates $\leq 1\%$. Compared to existing schemes that utilize readout resonators for dissipation~\cite{chen2024,yang2024}, our approach significantly shortens reset time and minimizes crosstalk in multi-qubit systems.

\begin{figure*}[t]
	\begin{center}
		\includegraphics[width=0.95\textwidth]{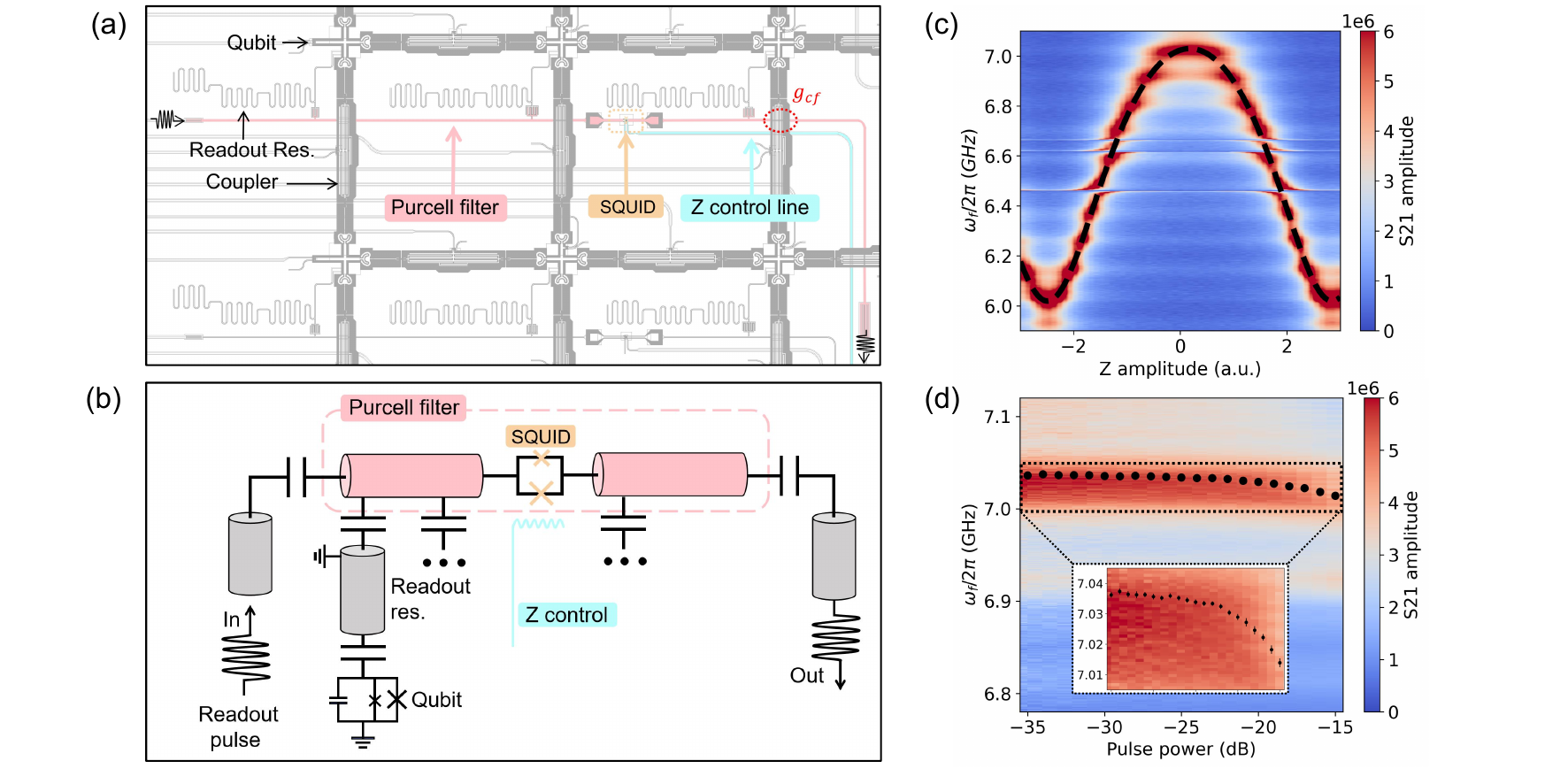}
		\caption{Tunable purcell filter integrated in a flip-chip quantum processor. (a) Filter-related structure. The filter (pink) consists of a half-wavelength resonator with open-ended capacitance coupling to the transmission line. The midpoint of filter is integrated with a SQUID (orange) containing two Josephson junctions, of which the flux is modulated by the Z control line(blue).  Each filter is shared by three readout resonators for qubit readout and coupled with couplers(circled in red) for qubit reset. (b) Distributed-element circuit model of the relevant device section with color coding matching (a). The ellipses denote other resonator-qubit systems and couplers capacitively coupled to the filter. (c) The variation of S21 amplitude under different Z amplitude applied to the SQUID. Black dashed curve indicates fitted resonance frequencies of filter passbands. (d) Power-dependent Kerr nonlinearity effect on the filter passband frequency. Black markers represent fitted resonance frequencies at the passband peaks.}
		\label{fig1}
	\end{center}
\end{figure*}

Our experiment employs a flip-chip superconducting quantum processor, interconnected via indium bump bonding. The device consists of aluminum-based coplanar waveguide circuits fabricated on sapphire substrates. The top chip contains 24 qubits, 38 couplers, and 8 Superconducting Quantum Interference Devices (SQUIDs) dedicated to filters, while the bottom chip serves as the wiring layer housing readout resonators, the main bodies of filters, and associated transmission lines. As shown in Fig.~\ref{fig1}(a)~and~(b), each tunable Purcell filter is parametrically coupled to three readout resonators via capacitance. A filter comprises a nonlinear structure, based on a $\lambda/2$ cavity with open ends capacitively coupled with transmission-line, incorporating a SQUID (composed of two Josephson junctions in parallel) at its midpoint. Note that the Josephson junctions of a filter are located on the top layer and are connected to the geometric cavity on the bottom layer through indium bumps. This design offers the advantage of enabling simultaneous fabrication of the Josephson junctions of filters, qubits, and couplers, thereby simplifying the fabrication process. Inputting DC signals through the Z-control line generates flux bias to tune the filter frequency among 6.02~GHz to 7.06~GHz (Fig.~\ref{fig1}(c)). 
The power-dependent Kerr nonlinearity effect is another reason for frequency changes. The nonlinear frequency shift remains negligible at low power levels (Fig.~\ref{fig1}(d)) because of a small anharmonicity of $\alpha_{f}/2\pi=4.47~\text{MHz}$. The frequency of this filter can be dynamically tuned through SQUID Z amplitude and readout pulse amplitude. This enables real-time optimization of effective linewidth $\kappa_{\text{eff}}$ of readout resonators through the filter, aiming to achieve optimal readout conditions (typically targeting $2\chi=\kappa_{\text{eff}}$ where $2\chi$ is the dispersive shift). This addresses the limitation of fixed-bandwidth linear filters, where fabrication-induced frequency deviations may cause  $2\chi\neq\kappa_{\text{eff}}$. And the tunable Purcell filter passband frequency expands the frequency allocation range of readout resonators on a single feedline, thereby mitigating frequency crowding in multi-qubit readout.

\begin{figure*}[htbp]
	\begin{center}
		\includegraphics[width=0.95\textwidth]{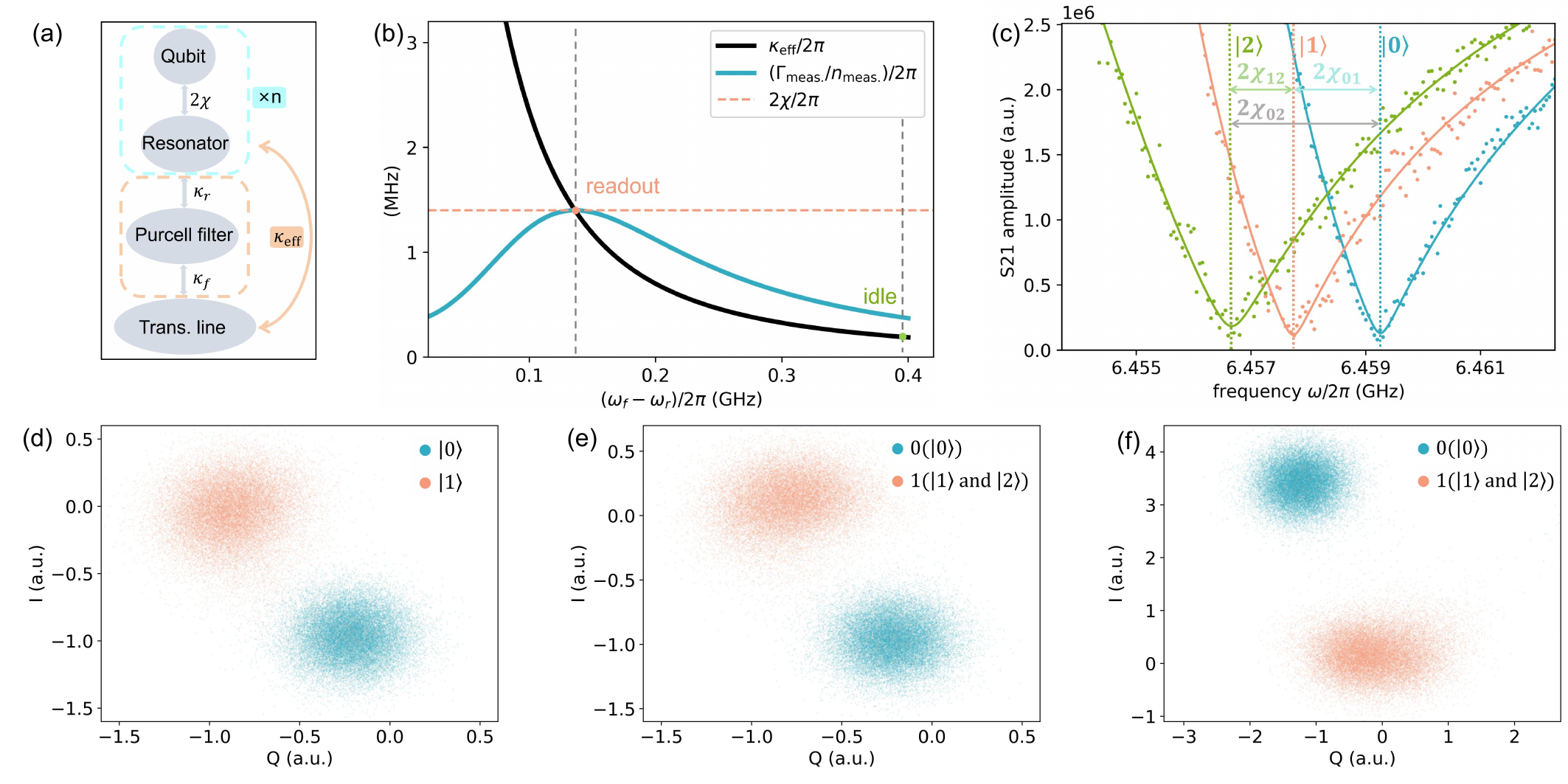}
		\caption{Experimental demonstration of flexible readout. (a) Schematic of the qubit readout system with tunable Purcell filters. The tunable Purcell filter functions to adjust $\kappa_{\text{eff}}$. Each filter is coupled with n resonators(n=3 here). (b) Variation curves of $\kappa_{\text{eff}}$ and $\Gamma_{\text{meas}}/n_{\text{meas}}$ versus frequency difference between the filter and readout resonator. The filter frequency resides at a point maximizing $\Gamma_{\text{meas}}/n_{\text{meas}}$  (where $\kappa_{\text{eff}}=2\chi$) during readout, while at a lower point during idle time. (c) The fitted dressed-state frequencies of the readout resonator with different qubit states. To maximize state distinguishability when distinguishing qubit states $\ket{i}$ and $\ket{j}$, the effective linewidth is set to $\kappa_{\text{eff}}=2\chi_{ij}$. (d-f) Typical data of IQ raw data: (d) $500~\text{ns}$ readout with direct measurement‌. (e) $500~\text{ns}$ readout with pre-$\pi_{12}$-pulse, mitigating qubit decoherence effects compared to (d)‌. (f) $1077~\text{ns}$ readout duration with pre-$\pi_{12}$-pulse, achieving enhanced SNR over (e) and $99.3\%$ readout fidelity.}
		\label{fig2}
	\end{center}
\end{figure*}
\begin{figure}[!htb]
	\includegraphics[scale=0.38]{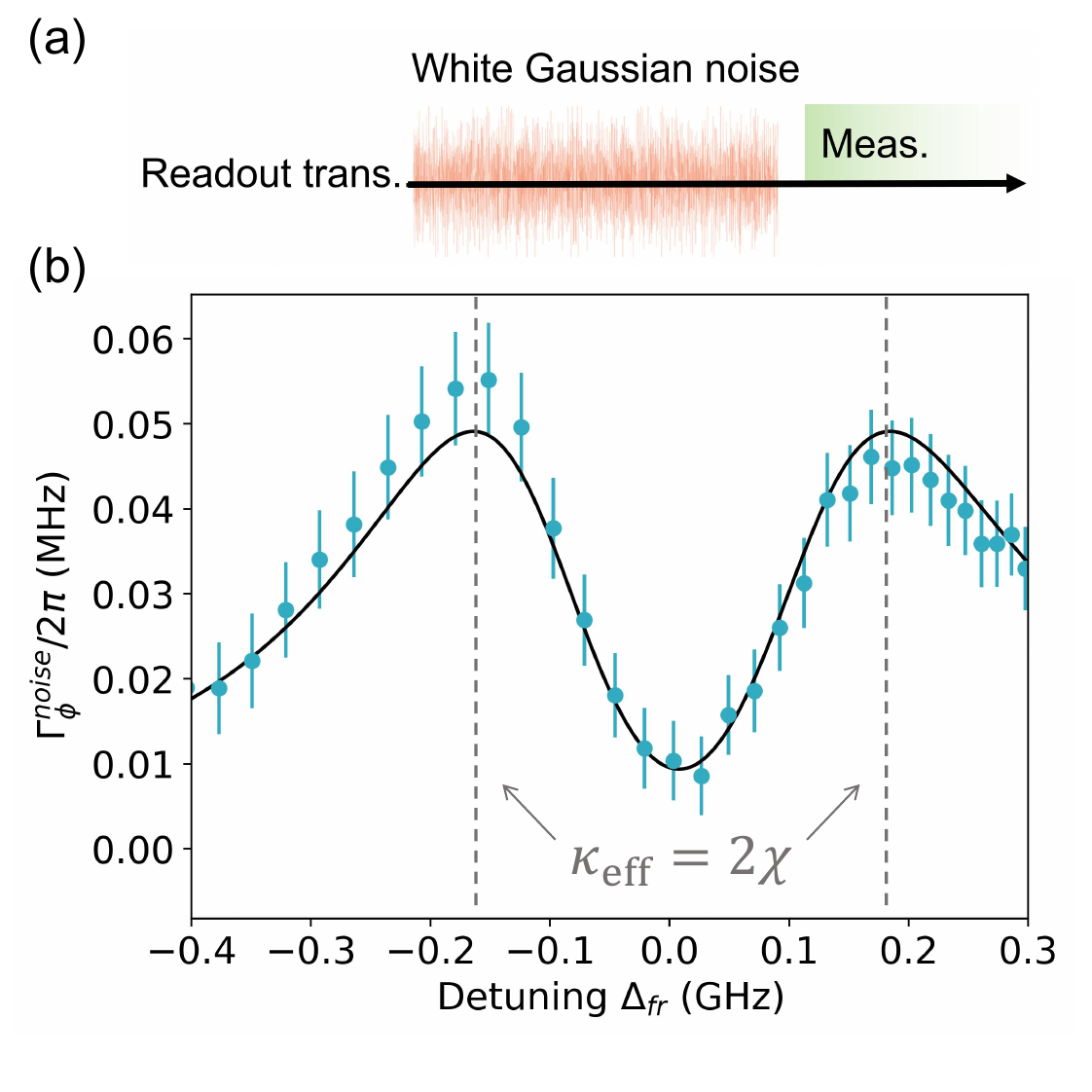}
	\caption{\label{fig3}Evaluation of the noise tolerance. (a) Pulse sequence.‌ The qubit $T_2$ is measured while applying white Gaussian noise through the readout transmission line. ‌(b) Qubit dephasing rate versus detuning. The dephasing rate $\Gamma_{\text{meas}}$ reaches a maximum when the detuning $\Delta_{fr}$ satisfies $\kappa_{\text{eff}}=2\chi$. The black curve shows the fit using a photon-noise-induced dephasing model (see Supplemental Material for details).
	}
\end{figure}

The SNR represents a crucial metric in high fidelity dispersive readout~\cite{blais2021,yufeng2024}, scaling with
$
\text{SNR}^2\propto \eta \kappa \bar{n} |\sin(2\theta)|^2t,
$
with $
|\sin(2\theta)|=\chi \kappa/(\chi ^2 +\kappa^2/4),
$
where $\kappa$ denotes the linewidth of the readout resonator, $\eta$ is the quantum efficiency, $\overline{n}$ is the average phton number, and $t$ is the readout time. $|\sin(2\theta)|^2=1$ is satisfied when $\kappa=2\chi$.
Figure~\ref{fig2}(a) illustrates the readout scheme incorporating a tunable Purcell filter, of which the role manifests in modifying $\kappa_{\text{eff}}$ to optimize SNR. Filter frequency modulation induces corresponding changes in $\kappa_{\text{eff}}$, as shown in Fig.~\ref{fig2}(b). The measurement-induced dephasing rate $\Gamma_{\text{meas}}$ is defined by $\Gamma_{\text{meas}}/n_{\text{meas}}=2\chi |\sin(2\theta)|$, where $n_{\text{meas}}$ is the average photon number in readout resonator. And $\Gamma_{\text{meas}}$ is the upper limit of measurement rate~\cite{clerk2010}. During active readout operations, the filter frequency aligns with the $\kappa_{\text{eff}}=2\chi$ configuration to maximize SNR and $\Gamma_{\text{meas}}/n_{\text{meas}}$ with a given $2\chi$. During idle periods, the filter frequency remains at a low-dephasing-rate operating point (idle point, where $\kappa\neq 2\chi$), enabling enhanced suppression of photon noise compared to conventional fixed-frequency filters. To verify the tunable Purcell filter's suppression effect on photon noise, we intentionally inject white Gaussian noise into the readout transmission line and measure the dephasing rate of a qubit, as shown in Fig.~\ref{fig3}.‌ 
Furthermore, the idle point maintains greater detuning between filter and qubit, providing additional suppression of Purcell effects. \par
Notably, the dispersive shift $2\chi$ of dressed states under different qubit states is not a fixed parameter (Fig.~\ref{fig2}(c)). This observation highlights important considerations: while direct measurements of qubit $\ket{0}$-$\ket{1}$ remain standard practice, contemporary protocols also  employ $\pi_{12}$-pulse-assisted $\ket{0}$-$\ket{2}$ readout~\cite{elder2020,gao2025}. $\ket{0}$-$\ket{2}$ readout means implementing a non-discriminating readout between $\ket{1}$ and $\ket{2}$ states, wherein: (i) All results measured in either $\ket{1}$ or $\ket{2}$ are registered as logical `1'. (ii) Only $\ket{0}$ measurements yield logical `0' outputs. The logical `0' or `1' is the single-shot measurement result. The filter provides adaptable $\kappa_{\text{eff}}$ values to accommodate various readout requirements. For instance, Figs.~\ref{fig2}(d) and (e) present IQ raw data for $\ket{0}$-$\ket{1}$ and $\ket{0}$-$\ket{2}$ readouts with 500 ns integration times, achieving SNR of 4.6 and 4.9 respectively. The $\ket{0}$-$\ket{2}$ readout mitigates $T_1$-limited fidelity constraints, enabling extension of integration time to $1077~\text{ns}$ (Fig.~\ref{fig2}(f)) with SNR improvement to 6.4 and a readout fidelity $\geq 99.3\%$. Here readout fidelity is defined as: $F=(P(0|0)+P(1|1))/2$, where $P(a|b)$ denotes the probalility that qubit is prepared in $\ket{b}$ state and then measured in $\ket{a}$ state. And SNR is calculated with $\text{SNR}=2|m_0-m_1|/(\sigma_0+\sigma_1)$~\cite{blais2021,peter2024}, where $m_{0(1)}$ and $\sigma_{0(1)}$ are the mean and standard deviration of the Gaussian fits to IQ data of $\ket{0}(\ket{1})$. Notably, the extended readout duration originates from the weak coupling design between readout resonators and qubits, implemented to preserve qubit coherence for complex multi-qubit gate operations. Each IQ distribution derives from 30,000 single-shot measurements.

\begin{figure*}[htbp]
	\begin{center}
		\includegraphics[width=0.95\textwidth]{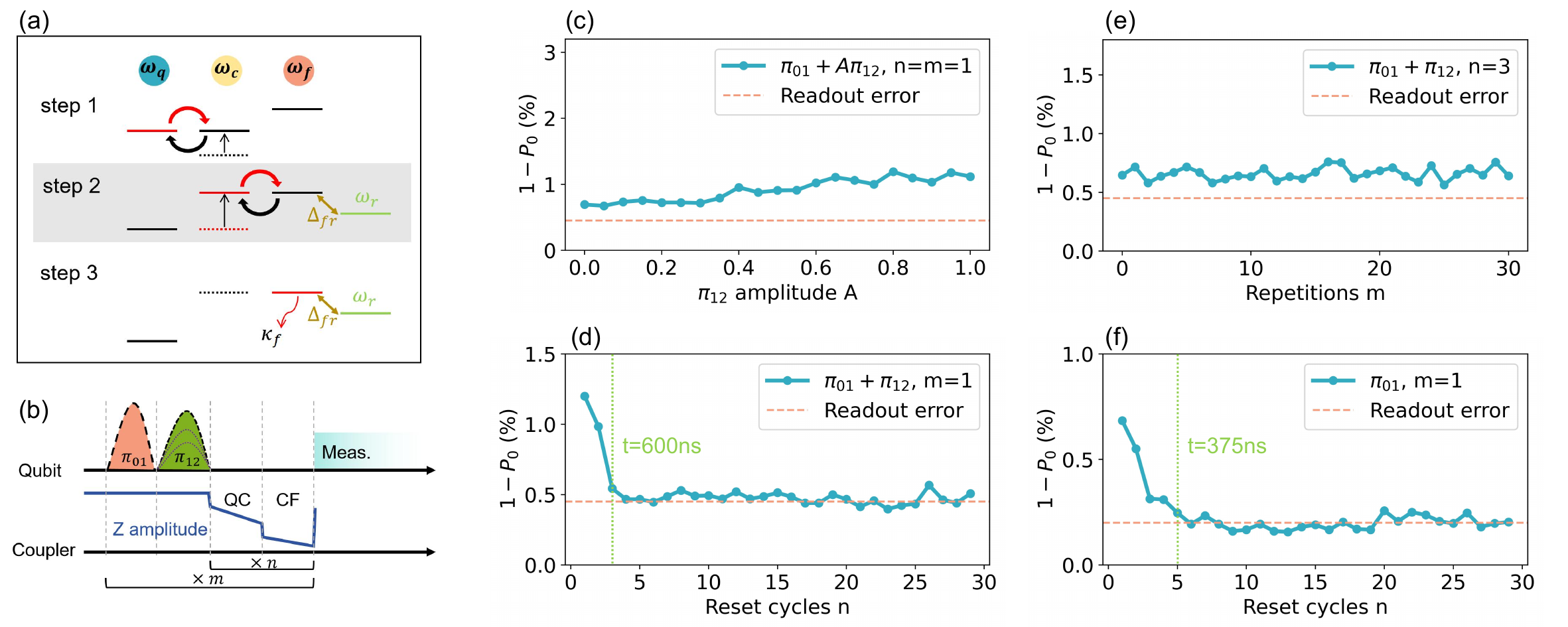}
		\caption{Experimental demonstration of unconditional qubit reset. (a) Reset protocol‌. ‌Step 1‌: Qubit-coupler (QC) swap. ‌Step 2‌: Coupler-filter (CF) swap. ‌Step 3‌: Dissipation of excitation into the environment via the Purcell filter. Steps 2 and 3 occur concurrently, with the filter detuned from the readout resonators to avoid crosstalk. ‌(b) Pulse sequence for state preparation, reset, and measurement. (c-f) Residual qubit excitation after reset, quantified by the measured $\ket{0}$ state population $P_0$. (c) Single reset cycle (200~ns), preparing initial state with $\pi_{01}$ and $A\pi_{12}$, where $0 \leq A\leq 1$. (d) Multiple reset cycles, with state preparation using $\pi_{01}$ and $\pi_{12}$. (e) Repeated state preparation (for $m$ rounds) followed by three reset cycles each time. (f) Multiple reset cycles with initialization using only the full $\pi_{01}$ pulse.}
		\label{fig4}
	\end{center}
\end{figure*}

Furthermore, we introduce a coupler-assisted qubit reset protocol leveraging the Purcell filter as a dissipation channel. Categorized into active (conditional)~\cite{rist2012,moreira2023} and passive (unconditional)~\cite{chen2024,zhou2021,wallraff2018} approaches based on whether qubit state information is required, rapid qubit reset techniques could be applied to various quantum information processing tasks especially the quantum error correction codes~\cite{google2023sqe}. While introducing additional decay channels risks complicating multi-qubit integration, the coupler-assisted reset offers a hardware-efficient solution‌~\cite{chen2024,yang2024}. However, existing schemes utilize the readout resonator for dissipation, whose rate is significantly lower than that of the filter. Employing the filter would further reduce the reset time and the error rate. In a flip-chip processor, qubits and couplers are typically placed on one layer, while transmission lines and Purcell filters are located on another layer~\cite{zhang2024,zhao2022,google2025}. Inevitably, filters traverse above couplers or qubits as shown in fig.~\ref{fig1}(a). The former results in a coupling strength $g_{cf}$ between filters and couplers. We ingeniously exploit this coupling to facilitate qubit reset. Crucially, the filter and readout resonators remain detuned throughout the process (filter: 7.0 GHz, readout resonators: 6.4--6.6 GHz in this experiment), effectively suppressing photon leakage into the readout resonators. This detuning strategy requires a frequency-tunable Purcell filter, as fixed-frequency filters must operate near the readout resonator frequency to maintain measurement fidelity. And the detuning avoid keeping the coupler at the readout resonator frequency, which would trigger complex multi-body processes.

As illustrated in Fig.~\ref{fig4}(a), our reset protocol comprises three sequential steps. ‌Step 1‌ involves adiabatic state swap between the qubit and coupler. The resonance between the coupler and the target qubit introduces non-zero coupling between the target qubit and another qubit connected via the same coupler. To ensure exclusive photon transfer to the coupler, the frequencies of these two qubits are intentionally detuned by 800 MHz here. ‌Step 2‌ utilizes the capacitive coupling mentioned above to perform adiabatic swap between the coupler and the filter.  ‌Step 3‌ involves rapid photon dissipation from the filter, enabled by its leakage rate (linewidth) $\kappa_{f}/2\pi=150~\text{MHz}$. It is noted that the linewidth of the filter is approximately 100 times larger than that of the readout resonator. Thus step 3 executes concurrently within operational timeframe of Step 2, requiring virtually no additional time. Especially, the coupler frequency starts below the target qubit frequency (and above the neighboring), sweeps through it, and continues ascending to the filter frequency to suppress Landau-Zener-Stückelberg interference~\cite{Kim2025fastunconditional}, which could otherwise degrade reset fidelity.

To validate the unconditional nature and leakage-error mitigation capability of this reset protocol, we conduct the experimental demonstration outlined in Fig.~\ref{fig4}(b). First, the qubit is prepared in varying excited states by applying $\pi_{01}$ pulses and $\pi_{12}$ fractional pulses of different amplitudes ($A\pi_{12}$ with $A$ from 0 to 1). Then a standardized reset operation (waveform details are shown in Supplemental Material) is executed by adjusting the coupler flux bias. State preparation is repeated for $m$ rounds. After each round, reset operations would be executed for n cycles. The result of a preparation round with $\pi_{01}+A\pi_{12}$ and a reset cycle is shown in Fig.~\ref{fig4}(c). The qubit-coupler swap and coupler-filter swap are completed within 30~ns and 170~ns, respectively, yielding a total reset within 200~ns. Immediate post-reset qubit measurements reveal the population of excited state below 1.2\% across all initial states. After accounting for a 0.45\% ground-state readout error (attributed to state discrimination limitation and natural thermal excitation), the unconditional reset error per cycle is inferred to below 1\%. 

Further investigation of consecutive multi-cycle reset operations after a round of state preparation with $\pi_{01}+\pi_{12}$, as shown in Fig.~\ref{fig4}(d), demonstrates progressive reduction of residual excitations with increasing reset cycles, converging to natural thermal excitation levels (quantified by a deviation less than $0.1\%$ from ground-state readout error line) after three cycles ($600~\text{ns}$). Critically, multiple reset cycles do not accumulate errors via photon recapture, confirming effective dissipation through the filter. Consider the following scenario: the qubit state is repeatedly prepared $m$ rounds with an interval of 1 microsecond, and the cycles of resets after each state preparation is $n=3$. As shown in Fig.~\ref{fig4}(d), the residual thermal excitation does not increase with the growth of $m$, which indicates that short-time multiple excitations and resets of the qubit is supported. In other words, this dissipation channel can be rapidly reused. When exclusively considering $\ket{1}$ excitation, we employ square pulse for QC and CF swap. The single-cycle reset operation can be optimized to complete within 75 ns (4.5 ns/ 70 ns for qubit-coupler/coupler-filter swap) with a error rate below 1\%. And the residual thermal excitation decreases and converges to the natural scenario after the 5-th cycle (375 ns), with a deviation from readout error line below 0.1\%. Each data point in Fig.~\ref{fig4}(c-f) is derived from 30,000 repetitions of single-shot measurements. The readout error line in Fig.~\ref{fig4}(f) differs from those in other figures dues to different readout parameters. Comparing with recent reports on the unconditional reset for quantum error correction~\cite{yang2024,google2023sqe,mcewen2021}, the time cost and error rate of our protocol are smaller. Therefore, our reset protocol has the potential to be applied in complex algorithms such as quantum error correction, where the reset operations are executed frequently. 

In conclusion, we present a scalable structure of tunable-frequency nonlinear Purcell filters for superconducting multi-qubit processors. Based on this structure, we implement a flexible readout protocol featuring the $\kappa_{\text{eff}}$-tuning capability of filters to maximize SNR during readout and suppress photon noise during idle time. Without quantum-limited amplifier, we achieve $99.3\%$ readout fidelity with a small dispersive shift of 1.4~MHz. By leveraging the coupling between couplers and filters in a flip-chip architecture, we propose an unconditional qubit reset protocol. This protocol eliminates both leakage-induced $\ket{2}$ and $\ket{1}$ states with a rate exceeds natural qubit relaxation by a factor of $10^3$. Thus, this tunable-frequency nonlinear Purcell filter structure represents a highly potential auxiliary hardware component to advance the development of fault-tolerant quantum computing systems. ‌We provide the parameter calculation, experimental detail, and numerical analysis in the supplementary material. Future improvements may focus on extending the filter geometry by utilizing the frequency-tunable property in a $n\lambda/2$ resonator to readout more qubits. Since the reset time is primarily limited by the coupler-filter coupling strength $g_{cf}$, increasing $g_{cf}$ could achieve faster reset operation.

This work is supported by the National Natural Science Foundation of China (Grants Nos. T2322030, T2121001, 92265207,  12122504, 12274142, 92365206, 12104055), the Innovation Program for Quantum Science and Technology (Grant No. 2021ZD0301800), the Beijing Nova Program (No. 20220484121), and the China Postdoctoral Science Foundation (Grant No. GZB20240815). We acknowledge the support from the Synergetic Extreme Condition User Facility (SECUF).

\bibliography{ref}  

\end{document}


\title{Supplemental Material for ``Flexible Readout and Unconditional Reset for Superconducting Multi-Qubit Processors with Tunable Purcell Filters''}

\author{Yong-Xi Xiao}
\thanks{These authors contributed equally to this work.}
\author{Da'er Feng}
\thanks{These authors contributed equally to this work.}
\author{Xu-Yang Gu}
\affiliation{Beijing National Laboratory for Condensed Matter Physics, Institute of Physics, Chinese Academy of Sciences, Beijing 100190, China}
\affiliation{School of Physics, University of Chinese Academy of Sciences Beijing 100049, China}

\author{Gui-Han Liang}
\affiliation{Beijing National Laboratory for Condensed Matter Physics, Institute of Physics, Chinese Academy of Sciences, Beijing 100190, China}
\affiliation{School of Physics, University of Chinese Academy of Sciences Beijing 100049, China}
\author{Ming-Chuan Wang}
\affiliation{Beijing National Laboratory for Condensed Matter Physics, Institute of Physics, Chinese Academy of Sciences, Beijing 100190, China}
\affiliation{School of Physics, University of Chinese Academy of Sciences Beijing 100049, China}
\author{Zheng-Yu Peng}
\affiliation{Beijing National Laboratory for Condensed Matter Physics, Institute of Physics, Chinese Academy of Sciences, Beijing 100190, China}
\affiliation{School of Physics, University of Chinese Academy of Sciences Beijing 100049, China}
\author{Bing-Jie Chen}
\affiliation{Beijing National Laboratory for Condensed Matter Physics, Institute of Physics, Chinese Academy of Sciences, Beijing 100190, China}
\affiliation{School of Physics, University of Chinese Academy of Sciences Beijing 100049, China}
\author{Yu Yan}
\affiliation{Beijing National Laboratory for Condensed Matter Physics, Institute of Physics, Chinese Academy of Sciences, Beijing 100190, China}
\affiliation{School of Physics, University of Chinese Academy of Sciences Beijing 100049, China}
\author{Zheng-Yang Mei}
\affiliation{Beijing National Laboratory for Condensed Matter Physics, Institute of Physics, Chinese Academy of Sciences, Beijing 100190, China}
\affiliation{School of Physics, University of Chinese Academy of Sciences Beijing 100049, China}
\author{Si-Lu Zhao}
\affiliation{Beijing National Laboratory for Condensed Matter Physics, Institute of Physics, Chinese Academy of Sciences, Beijing 100190, China}
\affiliation{School of Physics, University of Chinese Academy of Sciences Beijing 100049, China}
\author{Yi-Zhou Bu}
\affiliation{Beijing National Laboratory for Condensed Matter Physics, Institute of Physics, Chinese Academy of Sciences, Beijing 100190, China}
\affiliation{School of Physics, University of Chinese Academy of Sciences Beijing 100049, China}
\author{Cheng-Lin Deng}
\affiliation{Beijing National Laboratory for Condensed Matter Physics, Institute of Physics, Chinese Academy of Sciences, Beijing 100190, China}
\affiliation{School of Physics, University of Chinese Academy of Sciences Beijing 100049, China}

\author{Kai Yang}
\affiliation{Beijing National Laboratory for Condensed Matter Physics, Institute of Physics, Chinese Academy of Sciences, Beijing 100190, China}
\affiliation{School of Physics, University of Chinese Academy of Sciences Beijing 100049, China}    

\author{Ye Tian}
\affiliation{Beijing National Laboratory for Condensed Matter Physics, Institute of Physics, Chinese Academy of Sciences, Beijing 100190, China}

\author{Xiaohui Song}
\affiliation{Beijing National Laboratory for Condensed Matter Physics, Institute of Physics, Chinese Academy of Sciences, Beijing 100190, China}
\affiliation{School of Physics, University of Chinese Academy of Sciences Beijing 100049, China}

\author{Dongning Zheng}
\affiliation{Beijing National Laboratory for Condensed Matter Physics, Institute of Physics, Chinese Academy of Sciences, Beijing 100190, China}
\affiliation{School of Physics, University of Chinese Academy of Sciences Beijing 100049, China}
\affiliation{Beijing Academy of Quantum Information Sciences, 100193 Beijing, China}
\affiliation{Hefei National Laboratory, 230088 Hefei, China}
\author{Yu-Xiang Zhang}
\affiliation{Beijing National Laboratory for Condensed Matter Physics, Institute of Physics, Chinese Academy of Sciences, Beijing 100190, China}
\affiliation{School of Physics, University of Chinese Academy of Sciences Beijing 100049, China}

\author{Yun-Hao Shi}
\affiliation{Beijing National Laboratory for Condensed Matter Physics, Institute of Physics, Chinese Academy of Sciences, Beijing 100190, China}

\author{Zhongcheng Xiang}
\thanks{zcxiang@iphy.ac.cn}
\affiliation{Beijing National Laboratory for Condensed Matter Physics, Institute of Physics, Chinese Academy of Sciences, Beijing 100190, China}
\affiliation{School of Physics, University of Chinese Academy of Sciences Beijing 100049, China}
\affiliation{Beijing Academy of Quantum Information Sciences, 100193 Beijing, China}
\affiliation{Hefei National Laboratory, 230088 Hefei, China}

\author{Kai Xu}
\thanks{kaixu@iphy.ac.cn}
\affiliation{Beijing National Laboratory for Condensed Matter Physics, Institute of Physics, Chinese Academy of Sciences, Beijing 100190, China}
\affiliation{School of Physics, University of Chinese Academy of Sciences Beijing 100049, China}
\affiliation{Beijing Academy of Quantum Information Sciences, 100193 Beijing, China}
\affiliation{Hefei National Laboratory, 230088 Hefei, China}
\affiliation{Songshan Lake Materials Laboratory, 523808 Dongguan, Guangdong, China}

\author{Heng Fan}
\thanks{hfan@iphy.ac.cn}
\affiliation{Beijing National Laboratory for Condensed Matter Physics, Institute of Physics, Chinese Academy of Sciences, Beijing 100190, China}
\affiliation{School of Physics, University of Chinese Academy of Sciences Beijing 100049, China}
\affiliation{Beijing Academy of Quantum Information Sciences, 100193 Beijing, China}
\affiliation{Hefei National Laboratory, 230088 Hefei, China}
\affiliation{Songshan Lake Materials Laboratory, 523808 Dongguan, Guangdong, China}

\maketitle
\beginsupplement
\tableofcontents
\newpage

\section{\label{appB}SQUID parameters calculation for tunable filter }

\begin{figure}[!htb]
	\includegraphics[scale=0.45]{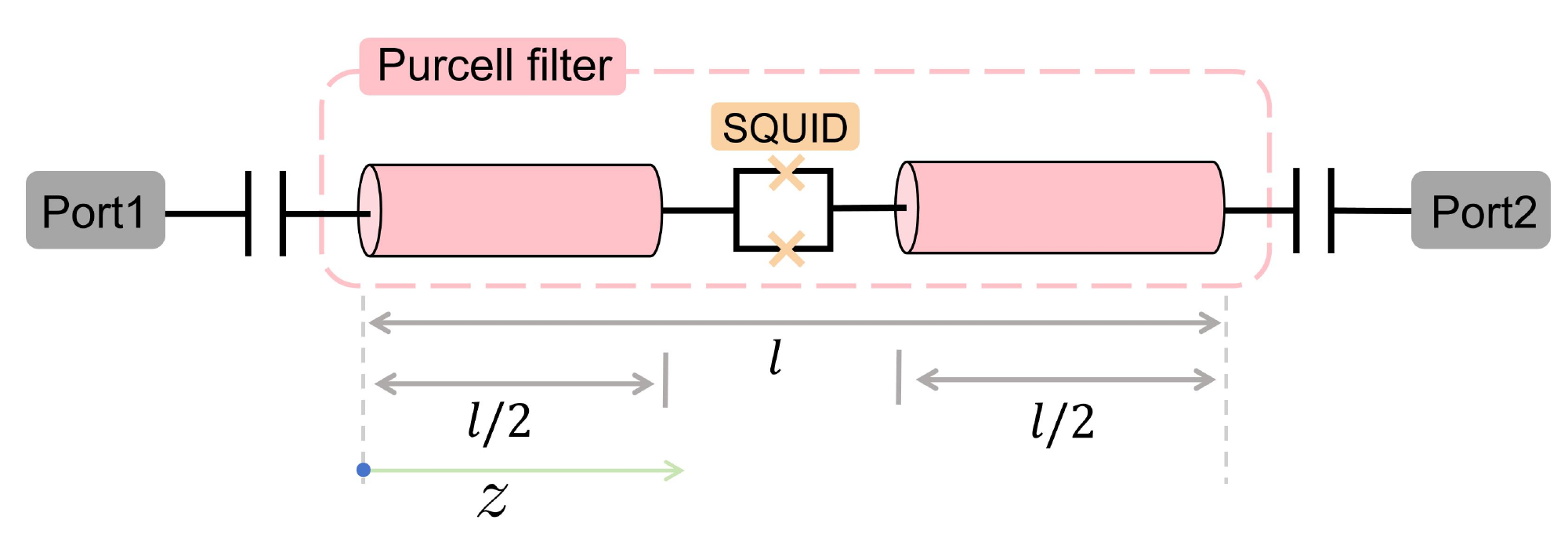}
	\caption{\label{figs1} Distributed-element circuit model of a tunable Purcell filter.}
\end{figure}

The filter is a $\lambda/2$ open-circuit cavity. A SQUID is embedded at the midpoint, enabling tunable cavity frequency by varying the magnetic flux (Fig.~\ref{figs1}).

First, consider the case without the SQUID: let $l$ denote the total cavity length, i.e., $l=\lambda/2$. The per-unit-length inductance and capacitance of the cavity are  $L_u$ and $C_u$, respectively, with a fundamental mode of $\omega_0=\pi /(l\sqrt{L_u C_u})$. For a position $z$ along the cavity (measured from the open end), the voltage and current are given by~\cite{2004Microwave}: 
\begin{equation}
	V(z)=V_{0}\cos (\beta_0 z),
\end{equation}
\begin{equation}
	I(z)=-\frac{jV_{0}}{Z_0}\sin (\beta_0 z),
\end{equation}
where $V_0$ is the voltage amplitude, $Z_0=\sqrt{L_u/C_u}$ is the characteristic impedance of the transmission line, $\beta_0=\omega_0/v_p$ is the wave vector, and $v_p=1/\sqrt{L_u C_u}$ is the phase velocity. At the midpoint $z=l/2$, the voltage is zero.
When a SQUID is inserted at the cavity midpoint $z=l/2$, the fundamental mode shifts to $\omega$. The SQUID is approximated as a point element that does not alter the cavity length. It introduces a voltage discontinuity at $z=l/2$, with voltages on either side given by:

\begin{equation}
	V_{l/2}^\pm =\pm V_{0}\cos( \frac{\beta l}{2}),
\end{equation}
where $\beta=\omega/v_p$ is the new resonant wave vector. The SQUID behaves as a superconducting inductance $L_S$ with impedance:

\begin{equation}
	Z_S=j\omega L_S.
\end{equation}
This impedance equals the ratio of the voltage difference across the SQUID to the current through it:

\begin{equation}
	\frac{2V_{0}\cos( {\frac{\beta l}{2}})}{-\frac{jV_{0}}{Z_0}\sin(  \frac{\beta l}{2})}=j\omega L_s,
\end{equation}
substituting $Z_0=\sqrt{L_u/C_u}$ and $\omega_0=\pi /(l\sqrt{L_u C_u})$, i.e., $2\pi Z_0 =2l\omega_0 L_u$ yields:

\begin{equation}
	\cot(\frac{\pi}{2}\frac{\omega}{\omega_0})=\frac{\pi  L_s}{2L_u l}\frac{\omega}{\omega_0}.
\end{equation}
Under the condition $(\omega_0 -\omega) \ll \omega_0$, we have 
\begin{equation}
	\cot(\frac{\beta l}{2}) \ll 1,
\end{equation}
allowing the approximation:

\begin{equation}
	\label{eqs8}
	\omega=\frac{\omega _0}{1+\frac{L_s}{L_u l}},
\end{equation}
 where $\omega_0=\pi /(\sqrt{L_{u}C_{u}}l)$, $L_{u}$ and $C_{u}$ represent the per-unit-length inductance and capacitance of the coplanar waveguide. This demonstrates the dependence of the filter's passband frequency on the applied magnetic flux bias. $L_S$ denotes the SQUID-induced inductance~\cite{oliver2019,blais2021}:

\begin{equation}
	\label{eqs9}
	L_s=\frac{\Phi_0}{2\pi\sqrt{(I_{c1}-I_{c2})^2+4I_{c1}I_{c2}\cos^2(\pi\Phi/\Phi_0)}},
\end{equation}
where $I_{c1}$ and $I_{c2}$ are the critical currents of the asymmetric junctions,  $\Phi$ is the flux and $\Phi_0=h/(2e)$. In this experiment, the filter frequency is primarily controlled by flux as Eq.~\ref{eqs8} and Eq.~\ref{eqs9}. Given the minimal anharmonicity of the filter, the frequency shift induced by readout pulse power variation remains relatively small, requiring only minor compensation through flux bias for correction. And the nonlinear physical properties related to the amplitude of the input field have been thoroughly described in previous studies~\cite{vijay2009,mallet2009,boaknin2007,sunada2024}, which will not be reiterated here.

\section{\label{appC}Readout circuits parameter with purcell filter}

The circuit design considers the respective parameters of the qubit, readout resonator, and filter, as well as their interaction. A general approach is as follows: \par First, a target effective readout resonator linewidth $\kappa_{\text{eff}}$ is specified based on readout time requirements, which simultaneously defines the matching readout resonator dispersive shift $2\chi=\kappa_{\text{eff}}$. The dispersive shift typically falls within the MHz range in multi-qubit devices to balance the readout SNR and the qubit protection. This target value is then converted to fundamental circuit parameters. Using the formula~\cite{sete2015}
\begin{equation}
	\kappa _{eff}=\frac{4|g_{rf}|^2}{\kappa _f}\frac{1}{1+\left( 2\Delta _{fr}/\kappa _f \right) ^2},
\end{equation}
we can relate the coupling strength $g_{rf}$ between readout resonator and filter, the frequency detuning $\Delta_{fd}=\omega_f-\omega_r$ between filter and resonator, and the filter's linewidth $\kappa_{f}$.
For systems dominated by capacitive coupling, the coupling strength between the readout resonator and the filter is given by
\begin{equation}
	g_{rf}=\frac{1}{2}C_{rf}\sqrt{\frac{\omega_r\omega_f}{C_rC_f}},
\end{equation}
where $C_{rf}$ represents the mutual capacitance between filter and readout resonator, while  $C_{f}$ and $C_{r}$ denote the equivalent ground capacitances at the coupling points of the filter and readout resonator, respectively. The filter linewidth is determined by
\begin{equation}
	\kappa_f=\frac{\omega_f}{Q_f},\ Q_f=\frac{C_f^{\text{\text{out}}}}{\omega_f Z_0 {C_{\text{out}}}^2}.
\end{equation}
Here, $C_{\text{out}}$ represents the coupling capacitance between filter output port and transmission line, and $C_f^{\text{out}}$ is the equivalent ground capacitance at the filter's output port. The condition $C_{\text{out}}\gg C_{in}$ is typically satisfied in S21 readout schemes to ensure most of the readout signal propagates towards the output port, which is essential for performing transmission readout with high measurement efficiency~\cite{bultink2018,blais2021}. Notably, for a $\lambda/2$ filter cavity in GHz frequency range (with mm-scale physical length), the spatial distribution of electric field must be carefully considered when determining equivalent ground capacitance at different positions. This can be efficiently analyzed using transmission line Y-parameters: when injecting a microwave signal at a specific position, the admittance parameter at that point satisfies~\cite{2004Microwave}
\begin{equation}
	Y_{11}=j2C(\omega-\omega_0),
\end{equation}
where the slope of the imaginary part of $Y11$ versus frequency $\omega$ directly yields twice the equivalent ground capacitance. This method enables accurate determination of capacitance values through simulation or experimental characterization.

Furthermore, the dispersive shifts are different for 01 and 02 measurements~\cite{koch2007,elder2020}. Here, we only consider the nearest-neighbor coupling. In dispersive regime, the dressed modes of a readout resonator are
\begin{equation}
	\omega _{r,0}=\omega _r-\chi _{0,1},
\end{equation}
\begin{equation}
	\omega _{r,1}=\omega _r+\chi _{0,1}-\chi _{1,2},
\end{equation}
\begin{equation}
	\omega _{r,2}=\omega _r+\chi _{1,2}-\chi _{2,3},
\end{equation}
where $\omega_{r,k}$ ($k=0,1,2$) is the dressed frequency with qubit at $\ket{k}$, $\omega_r$ is bare mode and $\chi _{i,j}$ is given by
\begin{equation}
	\chi _{i,j}=\frac{g_{i,j}^{2}}{\omega _{i\rightarrow j}-\omega _r}.
\end{equation}
Using the relation
\begin{equation}
	g_{j,j+1}\propto \sqrt{\frac{j+1}{2}}\left( \frac{E_J}{8E_C} \right) ^{1/4},
\end{equation}
the required dispersive shift can be expressed as
\begin{equation}
	\chi _{0,1}=\frac{g^2}{\omega _{0\rightarrow 1}-\omega _r},
\end{equation}
\begin{equation}
	\chi _{1,2}=\frac{2g^2}{\omega _{1\rightarrow 2}-\omega _r},
\end{equation}
\begin{equation}
	\chi _{2,3}=\frac{3g^2}{\omega _{2\rightarrow 3}-\omega _r},
\end{equation}
where the qubit-readout resonator coupling strength is
\begin{equation}
	g=\frac{1}{2}C_{qr}\sqrt{\frac{\omega_q\omega_r}{C_qC_r}}.
\end{equation}
Thus, the dispersive shifts for 01 and 02 measurement can be calculated as
\begin{equation}
	2\chi=\omega _{r,1}-\omega _{r,0},
\end{equation}
\begin{equation}
	2\chi^{\prime}=\omega _{r,2}-\omega _{r,0}.
\end{equation}
We have now formulated $\kappa_{\text{eff}}$ and $2\chi$ (for 01 and 02 measurement) in the context of circuit parameters. 

\section{\label{appD}Other device parameters}

\begin{figure}[t]
	\includegraphics[scale=0.45]{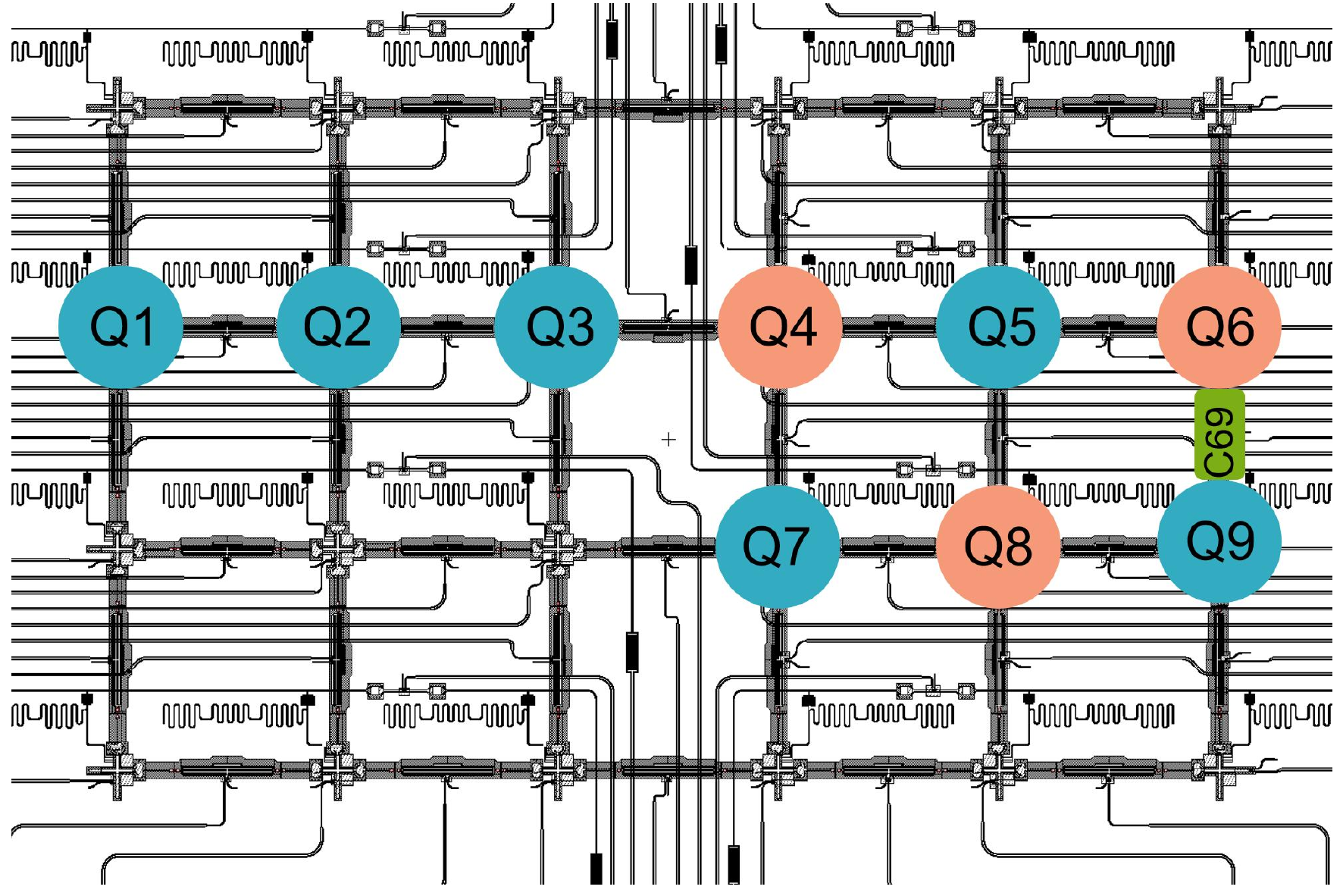}
	\caption{\label{figd1}Flip-chip sample diagram, where air bridges and indium bumps are intentionally omitted for clarity. The nine experimentally calibrated qubits with tunable frequencies are color-coded. Blue indicates qubits with a designed vertex frequency of 5.1 GHz, while orange represents those with a designed vertex frequency of 4.5 GHz. The reset scheme is implemented by $C_{69}$ (marked in green).}
\end{figure}

The experimental device is fabricated with a total of 24 qubits, employing the scalable coupling scheme proposed in Ref.~\cite{liang2023}. Considering the constraints of measurement and control equipment availability, we implement a 9-qubit subsystem (Fig.~\ref{figd1}) with tunable couplers for dynamic coupling strength modulation, enabling preliminary device characterization.

\begin{table}[ht!]
	\caption{\label{tableD1}Parameters of $Q_6$-$C_{69}$-$Q_9$ and average values across qubits $Q_1{\sim}Q_9$ on the flip-chip superconducting processor.}
	\begin{ruledtabular}
		\begin{tabular}{cccccc}
			& $Q_6$ & $C_{69}$ &$Q_9$ & $Q_{1}\sim{Q_9}$\\
			\hline
			$\omega^{\text{idle}}_{1,0}/2\pi$ (GHz)
			& 4.83 & 8.60 & 4.02 & 4.55  \\
			$\alpha/2\pi$ (MHz)
			& 188 & 120 & 193 & 190  \\
			$T_1^{\text{idle}}$ ($\mu {\rm s}$)
			& 35.6 &7.3 &48.3 &45.1  \\ 
			$T_2^{\text{CPMG}}$ ($\mu {\rm s}$)
			& 19.03 & &46.62 & 38.11\\ 
			$\omega_r/2\pi$ (GHz)
			& 6.45& &6.55 &6.59  \\
			$2\chi_{1,0}/2\pi$ (MHz)
			& 1.4& &1.1 & 1.2\\
			$F_{\text{meas.}}$ (\%)
			& 99.3& &98.1 & 96.7\\
			
		\end{tabular}
	\end{ruledtabular}
\end{table}

\begin{table}[h!]
	\caption{\label{tableD2}Parameters of tunable Purcell filter.}
	\begin{ruledtabular}
		\begin{tabular}{cccccc}
			& Parameter& ~ &Units\\
			\hline
			$C_{\text{in}}$ & Input capacitance & 15.7&fF\\
			$C_{\text{out}}$ & Output capacitance & 106.5&fF\\
			$I_c$ & Critical current of SQUID  & 0.4 &$\mu \text{A}$\\
			$I_{c1}$ & Critical current of Jose. junc. 1  & 0.085 &$\mu \text{A}$\\
			$I_{c2}$ & Critical current of Jose. junc. 2  & 0.315 &$\mu \text{A}$\\
			$\omega_0/2\pi$ & Frequency without SQUID & 9.3&GHz\\
			$\omega_{\text{bias}}/2\pi$& Frequency range tuned by flux bias & 6.02--7.06&GHz\\
			$\alpha_f/2\pi$ & Anharmonicity of filter & 4.47&MHz\\
			$\kappa_{f}/2\pi$& Linewidth of filter& 150  &MHz\\
			$L_u$ & Per-unit-length inductance & 464&nH\\
			$C_u$ & Per-unit-length capacitance & 203&pF\\
		\end{tabular}
	\end{ruledtabular}
\end{table}

To the impact of coupler-filter coupling capacitance on reset performance, we design various capacitive coupling strengths, with effective mutual capacitance values ranging from 0.64~fF to 2.4~fF. Considering different coupling positions on the filters, the coupling strength varies approximately from 4~MHz to 20~MHz. Except during reset, no significant effect of $g_{cf}/2\pi \leq 20~\text{MHz}$ on standard qubit gate operations is observed. This is because the filter remains strongly detuned from both qubits and couplers during non-reset operations. Here, $C_{69}$ represents the design with maximum coupling strength, enabling the fastest coupler reset. We focus on optimizing measurement and control parameters for $C_{69}$ and its two adjacent qubits $Q_6$ and $Q_9$. 
We designate $Q_6$ as an ancilla qubit and $Q_9$ as a data qubit. In surface code architectures, ancilla qubits are subject to repeated readout and reset during parity-check cycles, whereas data qubits remain largely protected from direct measurement-induced disturbances. Accordingly, the readout and reset performance reported in the main text is based on measurements of $Q_6$, while $Q_9$ is used to assess noise tolerance.

Device parameters used in our experiments are shown in Tab.~\ref{tableD1}. And the value of $2\chi$ obtained from the chi-kappa-power experiment~\cite{sank2024} is slightly different from that obtained by scanning S21 of dressed modes. Here we show the result from S21 scan. The readout fidelities of qubits other than $Q_6$ have not been fully optimized and could be further improved. The parameters of all filters on the device are nearly identical, as listed in Tab.~\ref{tableD2}. Note that, due to manufacturing tolerances, experimentally measured parameters may deviate from their design values. For instance, the actual measured $\kappa_f$ at filter frequency $\omega_f/(2\pi)\approx7~\rm{GHz}$ tends to be smaller than the value calculated. However, this discrepancy does not affect the overall performance. Taking the readout resonator of $Q_6$ as an example, the $\kappa_{\text{eff}}$ implemented by the filter is tunable between 0.2--11.5~MHz with a coupling of $g_{rf}/2\pi\approx20~\text{MHz}$. $I_c$ represents the Josephson junction critical current, which quantifies the nonlinearity introduced to the filter by the SQUID. Following Ref.~\cite{vijay2009}, the distributed-element circuit of the filter can be equivalently modeled as a lumped LCR circuit, thereby enabling the calculation of anharmonicity.

\section{\label{appE}Measurement and control setup}

\begin{figure}[!htb]
	
	\includegraphics[scale=0.5]{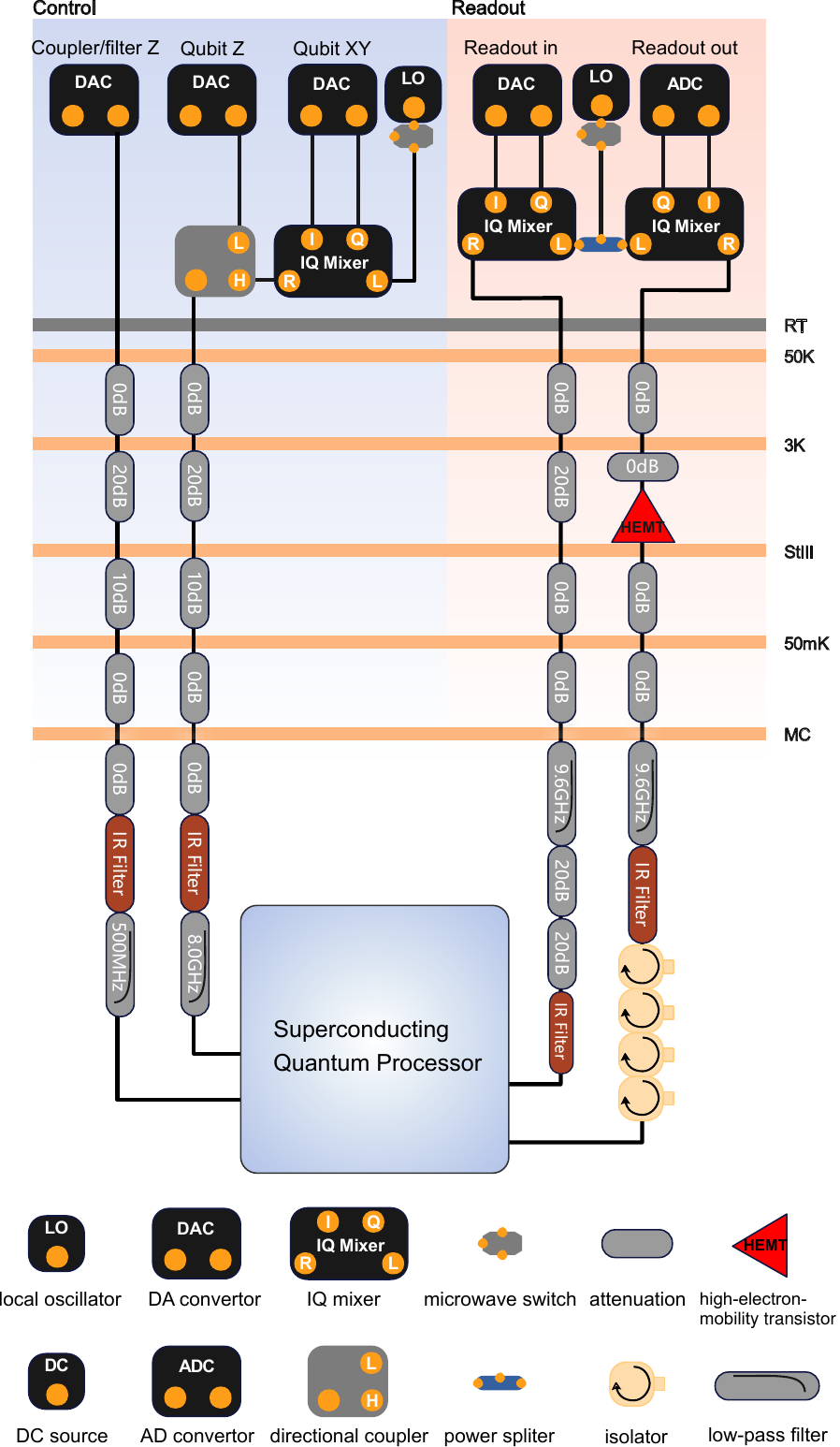}
	
	\caption{\label{fig:Cryogentic-and-room} Cryogentic and room temperature wiring}
\end{figure}
The quantum device is integrated into a BlueFors dilution refrigerator
(DR) system, with the mixing chamber plate (MC) stabilized at a base
temperature of approximately 7$\sim$13 mK. Standard cryogenic wiring
configurations are implemented for all qubit and tunable coupler
control lines as well as tunable filter flux control lines to ensure
optimal device performance. The configuration of the tunable filter
flux control lines is the same as that of the couplers. Comprehensive
microwave attenuation and filtering networks are installed to suppress
Johnson-Nyquist noise and higher-order harmonic interference. A multiplexed
readout architecture is employed, featuring a common feedline shared
among qubits, with four-stage isolation and a cryogenic low-noise
amplifier (LNA) for signal enhancement. Notably, there is no quantum-limited amplifier like JPA or TWPA.

The experimental sequence is orchestrated by a centralized control
system. All microwave control pulses are synthesized using a combination
of digital-to-analog converters (DACs) and upconversion through local
oscillator (LO) mixing.  The resulting quantum signals are subsequently
demodulated by LO signal and digitized by analog-to-digital converters
(ADCs) for data acquisition and analysis. A schematic representation
of the complete experimental setup is provided in Fig. \ref{fig:Cryogentic-and-room}.

\section{\label{appA}Suppression of photon noise and Purcell effect}

The photon-noise-induced dephasing rate is governed by~\cite{sunada2024}:

\begin{equation}\label{a1}
	\Gamma_{\phi}^{\text{noise}}=\frac{\kappa_{\text{eff}}(2\chi)^2}{\kappa_{\text{eff}}^2+(2\chi)^2}\overline{n}_{\text{noise}},
\end{equation}
where 
$\overline{n}_{\text{noise}}$
is the average photon noise number. When 
$\kappa_{\text{eff}}=2\chi$, the dephasing rate reaches its maximum $\Gamma_{\text{max}}$. Tunable filters maintain 
$\kappa\neq 2\chi$ during non-readout periods, ensuring 
$\Gamma_{\text{tunable}}<\Gamma_{\text{max}}$
, whereas fixed-frequency filters persistently operate at 
$\kappa=2\chi$ when qubit frequency is fixed, yielding 
$\Gamma_{\text{linear}}=\Gamma_{\text{max}}$. Thus, tunable nonlinear filters more effectively suppress qubit dephasing.
To verify the frequency-tunable filter's suppression effect on photon noise, we intentionally inject band-limited white Gaussian noise into the readout transmission line. The noise spectrum is distributed within a 1~GHz bandwidth centered on the readout frequency, with amplitudes following Gaussian distribution (mean = 0, standard deviation = 0.05). 
We then measure and fit the qubit dephasing rate.‌

‌Notably, the qubit in this experiment is frequency-tunable. We set the operating frequency of this qubit at a position below the vertex, which requires continuous flux bias to maintain. To better characterize how the dephasing time varies with white Gaussian noise intensity, we employed a Carr-Purcell-Meiboom-Gill (CPMG) sequence ($n=3$) to mitigate the effect of $1/f$-type noise.
\begin{figure}[!htb]
	\includegraphics[scale=0.4]{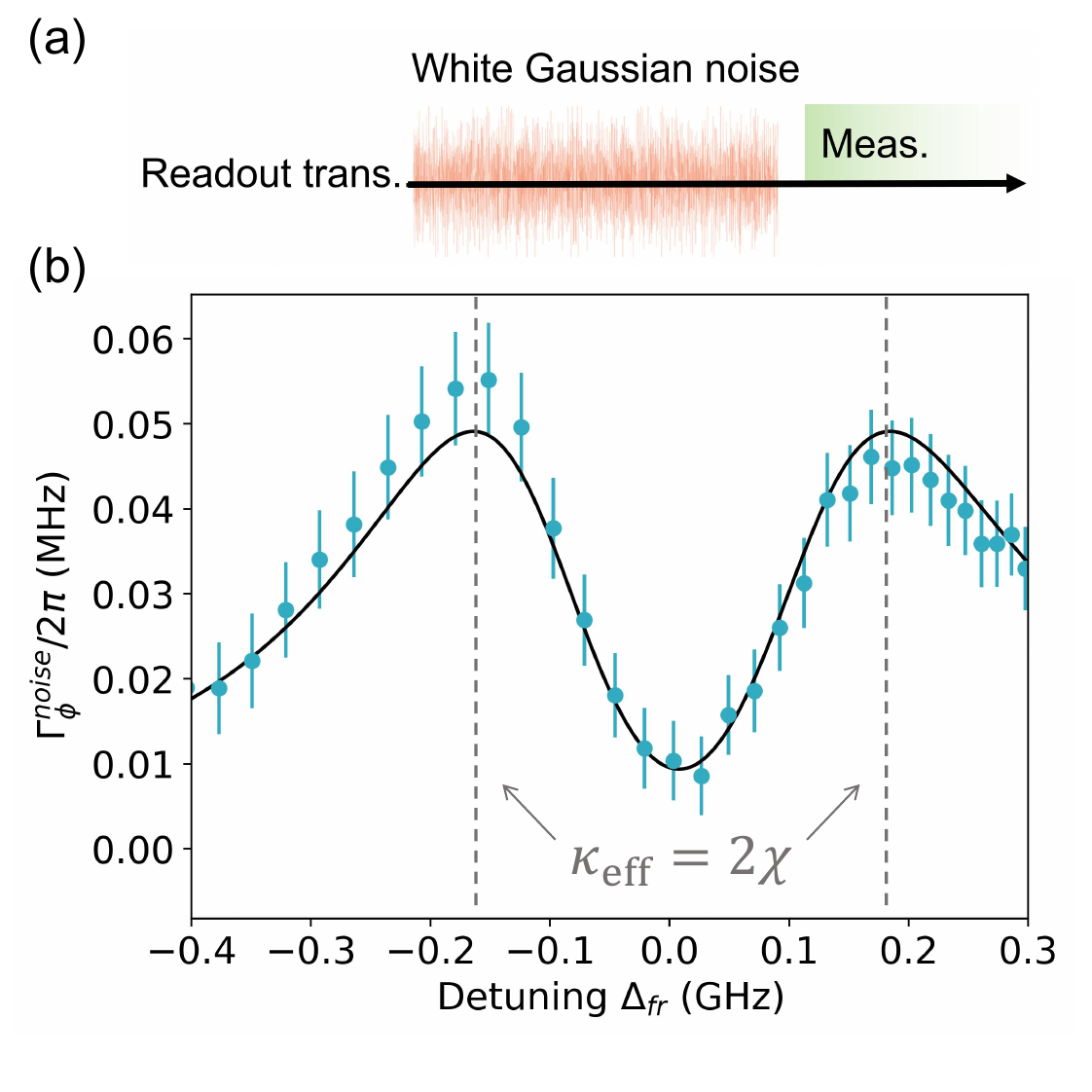}
	\caption{\label{fig4}Evaluation of the noise tolerance. (a) Readout pulse sequence.‌ The qubit $T_2$ is measured while applying white Gaussian noise through the readout transmission line. ‌(b) Qubit dephasing rate versus detuning.‌ The black curve represents a fit with Eq.~\eqref{a1}, where $\overline{n}_{\text{noise}}$ is regraded as a constant.}
\end{figure}

Measurement data under different  detunings of filter and readout resonator are shown in Fig.~\ref{fig4}, with error bars representing uncertainties in dephasing rate fitting. Error sources include residual flux bias and artificial noise amplitude fluctuations. The two peaks correspond to readout points at $\kappa_{\text{eff}}=2\chi$. Smaller detunings exhibit reduced dephasing rates due to excessive $\kappa_{\text{eff}}$, while larger detunings show decreased rates from insufficient $\kappa_{\text{eff}}$. Black curves in the figure represent fits using Eq.~\eqref{a1} assuming $\overline{n}_{\text{noise}}$ as a constant. The agreement between experimental data and theoretical predictions confirms the validity of photon noise suppression through frequency-tunable filters. Observed different peak height is attributed to photon absorption by other resonators sharing the same filter and increased transmission impedance at higher frequencies.

Tunable filter applications for multi-qubit dephasing suppression manifest in three aspects: First, the surface code scheme measures ancillary qubits to detect the occurrence of errors. The filter protects data qubits from measurement-induced dephasing if data qubits and ancillary qubits share the same readout line. Second, a protection window is created during post-measurement photon dissipation. Nonlinear filter frequency shift caused by decreasing photon numbers automatically reduces dephasing rate. Third, dephasing caused by non-ideal photon noise inputs is suppressed.
\par 
In the absence of a Purcell filter, the photon decay rate is expressed as:
\begin{equation}
	\Gamma=\kappa_{r}\frac{g_{qr}^2}{\left( \omega _r-\omega _q \right) ^2}.
\end{equation}
For samples incorporating a Purcell filter, the qubit photon decay rate induced by the Purcell effect is given by~\cite{man2015}
\begin{align}\label{a4}
	\Gamma _{01,\mathrm{\text{Purcell}}}
	&=\kappa_{\text{eff}}\frac{g_{qr}^2}{\left( \omega _r-\omega _q \right) ^2}\nonumber\\
	&=\frac{4|g_{fr}|^2\kappa _f}{\kappa _{f}^{2}+4\left( \omega _f-\omega _q \right) ^2}\frac{g_{qr}^2}{\left( \omega _r-\omega _q \right) ^2}.
\end{align}
The suppression of photon decay by fixed-frequency linear filters has been extensively discussed in prior studies. Here, we focus on comparing the difference between tunable nonlinear filter and fixed linear filter. For a fixed-frequency linear filter, the design typically requires $\omega_f \approx \omega_r$ with $\kappa=2\chi$. In contrast, a frequency-tunable nonlinear filter are not subject to such stringent constraints during non-readout periods, allowing a larger detuning from the qubit frequency. For instance, consider a scenario where both fixed-frequency filter and readout resonator operate at 6.5 GHz, while the tunable filter remains at 7 GHz (at idle point), with the qubit frequency at 5.5 GHz. The linewidth of both filters is the same $\kappa_{f}/2\pi =150~\text{MHz}$. The result derived from Eq.~\eqref{a4} yields 
\begin{equation}
	\frac{\Gamma_{\text{fixed,Purcell}}}{\Gamma_{\text{tunable,Purcell}}}\approx 2.08,
\end{equation}
which demonstrates that the larger detuning enabled by a tunable filter provides enhanced suppression of the Purcell effect compared to a fixed-frequency filter.

\section{\label{appF}Readout optimization procedure}

The tunable Purcell filter introduces an extra degree of freedom for
tuning $\kappa_{\mathrm{eff}}$ to achieve $2\chi\approx\kappa_{\mathrm{eff}}$,
and it also provides enhancement on readout fidelity via bifurcation
due to its nonlinearity. The readout optimization procedure is much
more complex since the flux bias voltage of tunable Purcell filter,
the resonator drive (readout) frequency and the resonator drive (readout) power are highly
correlated. For convertional fixed-frequency linear Purcell filter,
performing parameter scans with respect to resonator drive frequency
and power would usually make for the majority of improvement on readout
fidelity. However, with tunable Purcell filter, the frequency of which
is highly detuned from the resonators when no flux bias (Z pulse) is applied,
we need to jointly optimize flux bias with these parameters, and the
frequency of the Purcell filter would significantly affect readout
SNR. As performing parameter scan for more than two dimensions is
time consuming, we can opt to alternately sweep different two of
the parameter dimensions and iteratively converge toward a near-optimal
parameter set. 

\begin{figure*}[b]
	\begin{center}
		
		\includegraphics[scale=0.55]{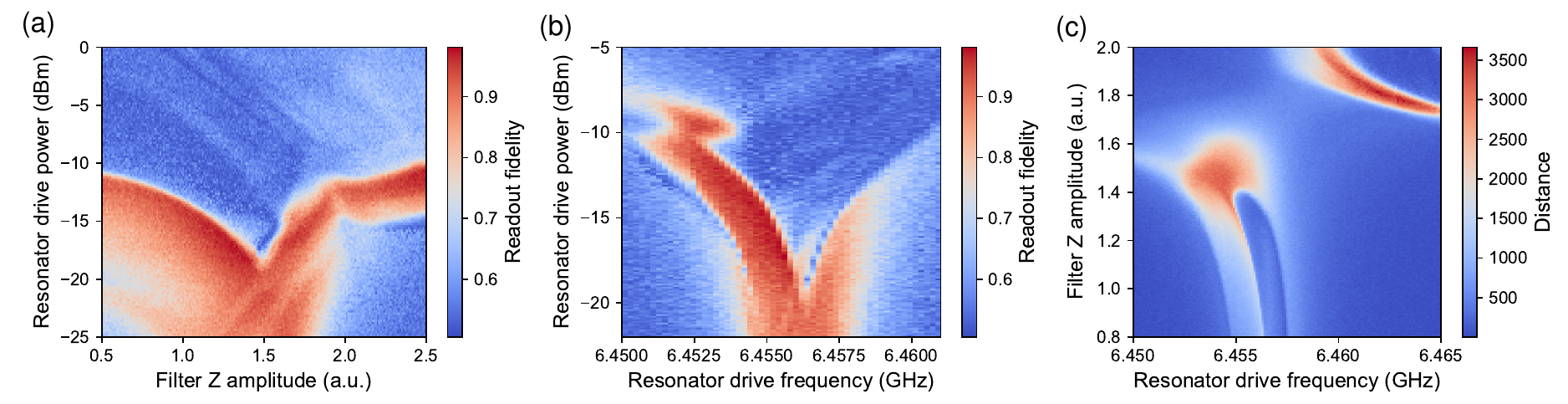}
		
		\caption{The relation of resonator drive frequency, power and filter Z pulse amplitude to readout fidelity or distance between readout signal
			centers of $|0\rangle$ and $|1\rangle$. (a) Readout fidelity under
			different resonator drive power and frequency with filter
			Z pulse amplitude being $1.32$. (b) Readout fidelity under different resonator
			drive power and filter Z pulse amplitude with resonator drive
			frequency being $6.4554$ GHz. (c) Distance between average values
			of readout signal of state $|0\rangle$ and $|1\rangle$ with resonator
			drive power $-17.4$~dBm.}\label{fig:readout_fid_scan}
	\end{center}
	
\end{figure*}

For illustration, we fix one of the parameters and scan a board range
of the other two parameters. The relationship between these parameters
and readout fidelity is shown in Fig.~\ref{fig:readout_fid_scan}.
As the power increases, bistability of the the resonator field emerges,
mainly due to high-order nonlinear terms of the qubit-resonator interaction,
consistent with the phenomenon reported in ref.~\cite{choi2024measurementinduced}.
Consequently, at the same readout power level, two distinct frequency
regions are available for achieving relatively high readout fidelity,
as is shown in Fig.~\ref{fig:readout_fid_scan}(a). The lower-frequency
region exhibits superior readout fidelity, primarily attributed to
its deeper dip characteristic. The two regions of Z pulse amplitude to
reach high readout fidelity at high readout power in Fig.~\ref{fig:readout_fid_scan}(b)
can also be explained by the bistability of the resonator field, where
the conditions $2\chi=\kappa_{{\rm eff}}$ for the two dips are reached
for two different filter Z pulse amplitudes respectively. The two resonator 
dips shift to diverse directions when readout power is increased, and thus it requires
to compensate this frequency shift by tuning the filter's frequency.
For low readout power, however, the bistability varnishes, thus there is only
one region for optimal readout fidelity under certain readout power.
As the frequency of the tunable Purcell filter get close to the resonator
frequency, anticross pattern of the resonator-filter system occurs,
and there are two frequency branches that can reach high readout fidelity,
as depicted in Fig.~\ref{fig:readout_fid_scan}(c).

\begin{figure}[!htb]
	
	\includegraphics[scale=0.6]{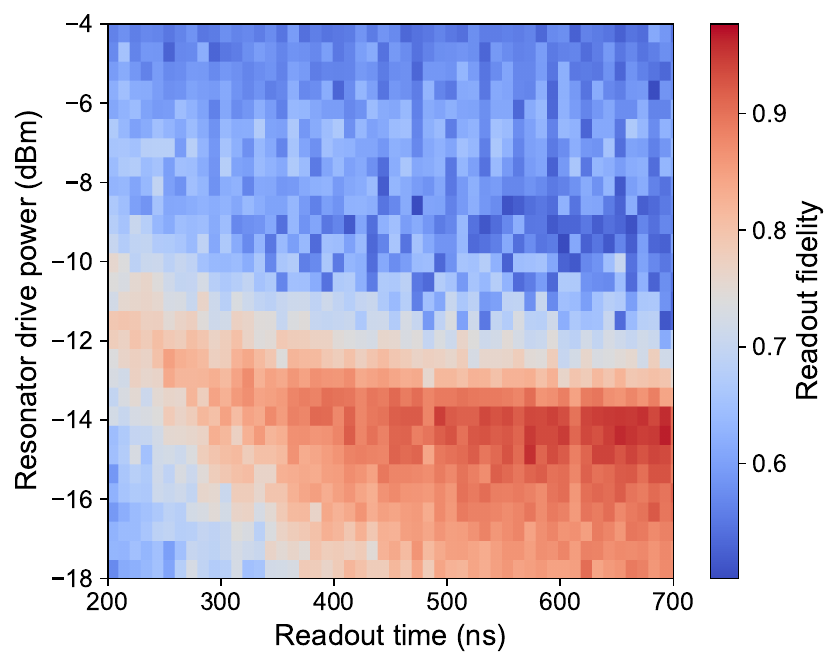}
	
	\caption{Parameter scan of readout fidelity with respect to readout time and
		resonator drive power.}\label{fig:readout-time-and-power}
	
\end{figure}

The readout time also play a crucial role in readout fidelity, as
we need to strike a balance between high SNR and low qubit state transition
rate during the readout process, and lowering readout time consumption
allows more operations under the limited decoherence time, especially
in scenarios where readout is applied multiple times such as quantum
error corrections. Parameter scan result for readout time and power
is presented in Fig.~\ref{fig:readout-time-and-power}. When the readout time is reduced, high readout fidelity can still be reached by slightly increasing readout power. Additionaly,
applying ring-up pulse and dynamially tuning qubit flux control can
improve SNR and reduce state transition rate, respectively. In such
scenario requiring simultaneous optimization of multiple variables,
employing grid search (parameter scan) inevitably explores extensive
low-fidelity parameter subspaces. Beyond conventional coarse-to-fine
scanning with alternating iterative parameter optimization for each
paramters, Bayesian optimization offers automated parameter exploration.
We utilize the Tree-structured Parzen Estimator algorithm from the
Optuna package ~\cite{akiba2019optunaa}, which can operate within
predefined parameter ranges to autonomously identify relatively optimal
hyperparameter configurations. Compared to manual iterative optimization,
heuristic approaches, and traditional gradient-free algorithms, its
advantage lies in the capacity to explore multiple high-potential
regions while avoiding persistent entrapment in local minima. This
methodology discovers multiple distinct relatively optimal parameter
combinations, which is explainable since we observe several regions
with comparatively high fidelity in the preceding parameter scan.

\begin{table}
	\caption{Selected optimal readout parameters for 500ns and 1077ns readout.}\label{tab:Selected-optimal-readout}
	
	\centering{}%
	\begin{tabular}{ccc}
		\hline 
		\hline
		Readout time (ns) & 500  & 1077\tabularnewline
		Filter readout ZPA & 0.680367 & 1.576001\tabularnewline
		Qubit readout ZPA & 0.440191 & 0.035742\tabularnewline
		Readout pulse power (dBm) & -1.0349 & -16.9254\tabularnewline
		Ring-up pulse power (dBm) & 6.5958 & -22.5849\tabularnewline
		Ring-up pulse time length (ns) & 12.0 & 65.0\tabularnewline
		Readout frequency (GHz) & 6.452666 & 6.453746 \tabularnewline
		\hline 
		\hline 
	\end{tabular}
\end{table}

\begin{table}
	\caption{Error budget analysis of qubit readout with optimal parameters for
		500ns $|0\rangle$-versus-$|1\rangle$ and $|0\rangle$-versus-$|2\rangle$
		readout and 1077ns $|0\rangle$-versus-$|2\rangle$ readout with post
		selection.}\label{tab:Error-budget-analysis}
	
	\begin{ruledtabular}
		\begin{tabular}{cccc}

			& 500 ns (01) & 500 ns (02) & 1077 ns (02) 
			\tabularnewline
			\hline 
			readout fidelity & 97.37\% & 98.03\% & 99.3\%\tabularnewline
			state $|0\rangle$ fidelity & 98.95\% & 99.36\% & 99.65\%\tabularnewline
			state $|1\rangle$ fidelity & 95.78\% & 96.70\% & 98.96\%\tabularnewline
			separation error & 0.147\% & 0.038\% & 0.0132\%\tabularnewline
			state error & 2.485\% & 1.929\% & 0.682\%\tabularnewline
			energy relaxation & 1.782\% & $\leq$1.782\% & $\leq$3.798\%\tabularnewline
			
		\end{tabular}
		\par\end{ruledtabular}
\end{table}

We select readout pulse length of $500$ ns and allow the readout time
to vary within a 2000 ns, respectively, and search for optimal readout
parameters via Bayesian optimization. For the latter, we identify
1077 ns as the pulse duration yielding superior readout fidelity through
optimization. We find optimal filter readout Z pulse amplitude (ZPA) from two different
regions, indicating both of the branches can reach high readout fidelity.
The selected parameters and the corresponding error budgets are listed
in Table~\ref{tab:Selected-optimal-readout} and ~\ref{tab:Error-budget-analysis}.
The ring-up pulse consists of a Gaussian envelope with 75~ns full-width
at half-maximum (FWHM), symmetrically truncated with respect to its central axis within the ring pulse time length interval, which is superimposed at the begining of the smoothed square readout pulse with a duration of the total readout length and 10ns FWHM Gaussian smoothing. The readout power refers to the output radio-frequency signal power level from
the room-temperature electronic devices. The readout ZPA denotes the amplitude of the
Z pulse applied during readout, maintaining temporal alignment with
the readout pulse duration. The separation error quantifies the misclassification
arising from the overlap region of the approximately Gaussian distributions
characterizing the readout IQ data for the prepared $\ket{0}$
and $\ket{1}$ (or $\ket{2}$) states. Conversely, the state
error represents the residual misclassification rate, primarily originating
from state transitions during readout and state preparation errors
such as imperfect $\pi$ pulse and residual excitation. A portion
of the transition error stems from energy relaxation, which can be
estimated by $\epsilon_{T_{1}}=1-\exp(-\tau_{m}/T_{1})$ with the
readout time $\tau_{m}$ and the calibrated $T_{1}\approx27.8~{\rm \mu s}$
of state $|1\rangle$ relaxation at current qubit frequency. For $|0\rangle$-versus-$|2\rangle$
readout, however, the relaxation-induced error rate from the $|2\rangle$
state to the $|1\rangle$ state remains uncalibrated. While the state
is initially prepared in state $|2\rangle$, it takes two stages of
relaxation to transition to state $|0\rangle$, i.e., $|2\rangle\rightarrow|1\rangle$
and $|1\rangle\rightarrow|0\rangle$; consequently, energy relaxation
errors are upper bounded by the state $|1\rangle$ relaxation error.
Furthermore, the misclassification rate depends critically on the
placement of the discrimination threshold. Rather than selecting the
perpendicular bisector between the $|0\rangle$ and $|1\rangle$ state
IQ centroids, we optimize readout fidelity by deliberately offsetting
the threshold toward state $|0\rangle$. This compensates for fidelity
loss caused by relaxations which pulls and distorts the distribution
of $|1\rangle$ state data points towards $|0\rangle$'s centroid.
While this threshold optimization enhances average readout fidelity,
it unavoidably lowers the  $|0\rangle$-state fidelity. This explains
why the $|0\rangle$-state fidelity may remain suboptimal despite
its low state transition rate and minimal separation error contribution. 

\begin{figure}[!htb]
	
	\includegraphics[scale=0.12]{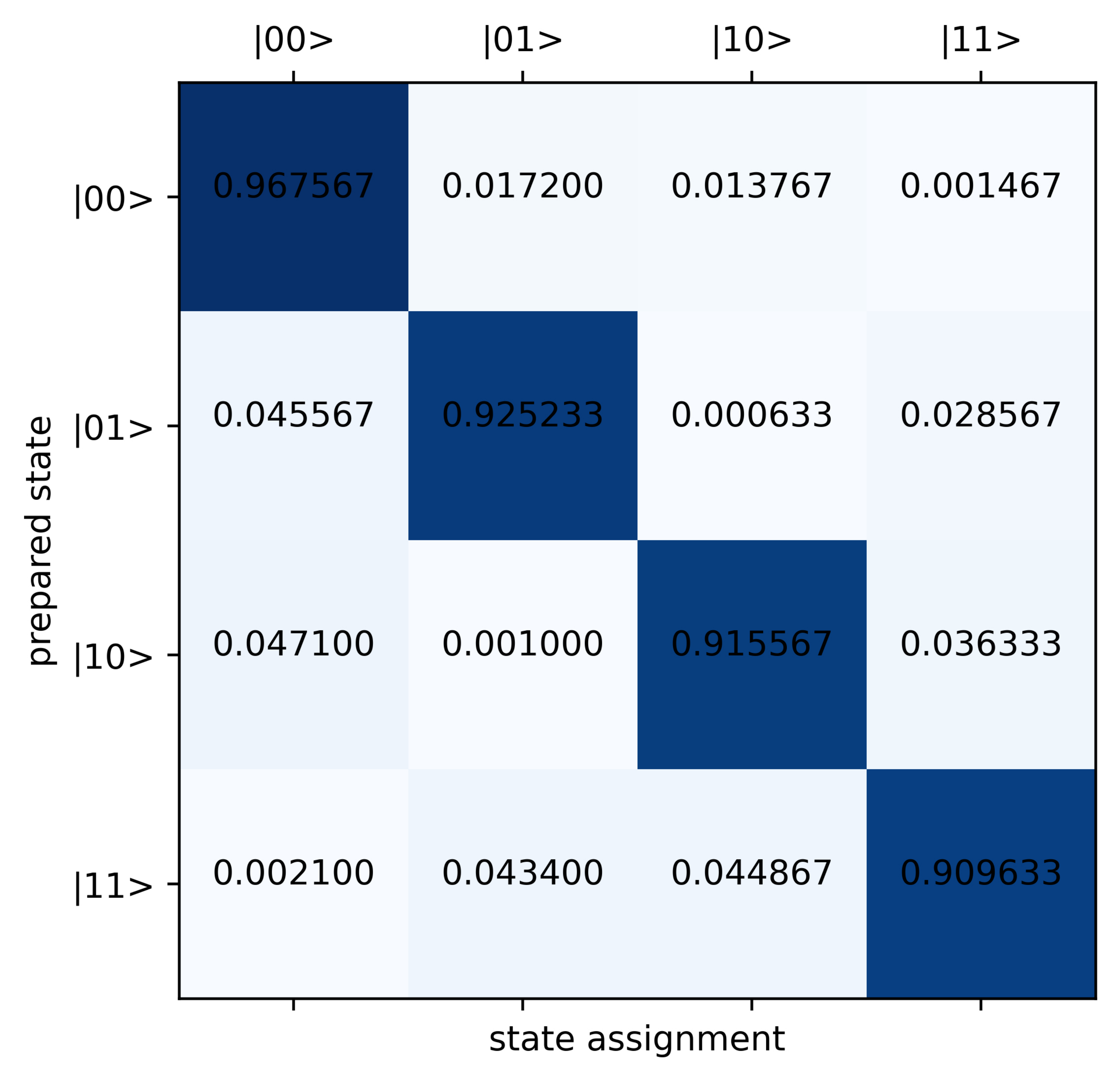}
	
	\caption{Joint readout fidelity of $Q_4$ and $Q_6$ from 30000 times single shot.}\label{figf3}
	
\end{figure}

Furthermore, we demonstrate multiplexed joint dispersive readout of multi-qubits by simultaneously applying multi-frequency readout pulses and then demodulating the acquired signals to extract measured qubit states respectively. Limited by Purcell filter linewidth $\kappa_{f}$, this experiment only achieved optimized results for two-qubit joint readout, as shown in Fig.~\ref{figf3}. The result indicates that tunable Purcell filter supports high-fidelity joint readout. Future extensions to high-fidelity joint readout of more qubits would require either increasing $\kappa_{f}$ or incorporating broadband filter designs~\cite{yan2023broad}.

\section{\label{appG}Reset experiment details}

The procedure of finding parameters for qubit reset is similar to
ref.~\cite{chen2024}. We adopt flat square pulse to perform qubit-coupler swap, and subsequently use flat square pulse to perform coupler-filter swap for state $|1\rangle$ reset. While for unconditionally resetting both $|1\rangle$ and $|2\rangle$, we implement adiabatic pulse for qubit-coupler swap, and a ramp pulse for coupler-filter swap. 

The flat pulses enable direct on-resonance exchange of a single excitation
from qubit to coupler and then from coupler to the Purcell filter,
which is simple to implement and fast for resetting only state $|1\rangle$.
While for unconditionally resetting both the $|1\rangle$ and $|2\rangle$
excitations of the qubit, the resonance exchange method is inadequate
for achieving complete transfer of both the $|1\rangle$ and $|2\rangle$
excitations from qubit to coupler in one qubit-coupler swap stage,
whereas the adiabatic exchange method is more preferable. Before the
reset operation, the qubit state is in arbitrary superposition of
$|1\rangle$ and $|2\rangle$. And then we adiabatically tune the
coupler frequency from below to above the qubit frequency, during
which the eigenstates of the interacting qubit-coupler system
exhibit negligible probability transfer. In the eigenbasis of the
non-interacting qubit and coupler Hamiltonian, two of these
eigenstates predominately overlap with $|1_{q}0_{c}\rangle$ and $|2_{q}0_{c}\rangle$
before the exchange, and predominately overlap with $|0_{q}1_{c}\rangle$
and $|0_{q}2_{c}\rangle$ after the exchange, respectively, thereby
accomplishing the simultaneous population transfer of both $|1\rangle$
and $|2\rangle$ states from the qubit to the coupler. The same adiabatic
protocol can applied to the coupler-filter swap stage as well.

\begin{figure}[!htb]
	\includegraphics[scale=0.55]{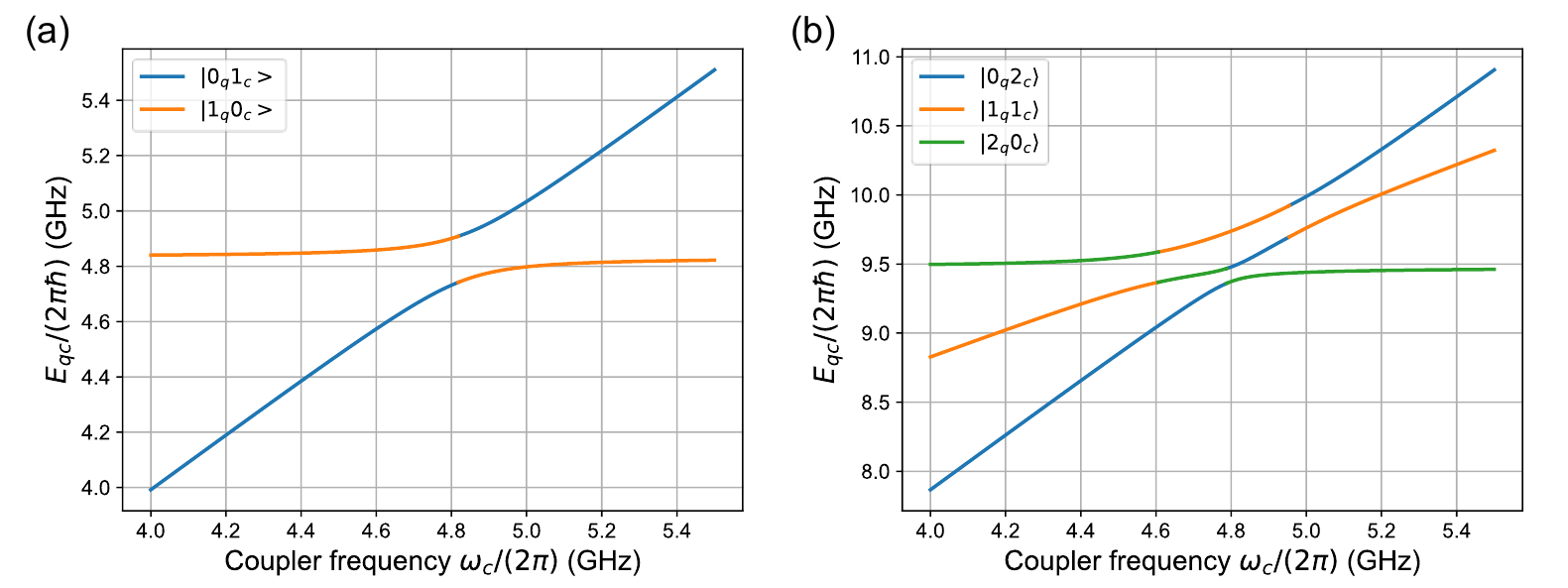}
	\caption{Energy spectrum of qubit-coupler Hamiltonian. The qubit frequency is set to $4.835~{\rm GHz}$, while the coupler's frequency $\omega_c/(2\pi)$ varies from $4.0~{\rm GHz}$ to $5.5~{\rm GHz}$. (a) Energy of eigenstates in $\{|0_q 1_c\rangle, |1_q 0_c\rangle\}$ subspace. (b) Energy of eigenstates in $\{|0_q 2_c\rangle, |1_q 1_c\rangle, |2_q 0_c\rangle\}$ subspace.}\label{fig:qcanticross}
\end{figure}

The adiabatic qubit-coupler swap pulse shape we use is derived in
~\cite{chen2024}, which is similar to the Roland
and Cerf protocol. For a qubit-coupler system with Hamiltonian 
\begin{equation}
	H_{qc}^{(1)}(t)=\hbar\begin{bmatrix}\Delta(t)/2 & g_{qc}\\
		g_{qc} & -\Delta(t)/2
	\end{bmatrix}
\end{equation}
in $\{|0_{q}1_{c}\rangle,|1_{q}0_{c}\rangle\}$ subspace, where $g_{qc}$
is the qubit-coupler coupling strength, and $\Delta(t)=\omega_{c}(t)-\omega_{q}$
is their $|0\rangle\leftrightarrow|1\rangle$ transition frequency detuning.
To satisfy adiabatic condition, a small scaling prefactor $\beta\ll1$
is introduced, such that 
\begin{equation}
	|\partial_{t}\Delta(t)|=\beta\left(\Delta^{2}(t)+4g_{qc}^{2}\right)^{\frac{3}{2}}/g_{qc}.\label{eq:adiabatic_condition_01anticross}
\end{equation}
While considering the reset of state $|2\rangle$, the qubit-coupler
Hamiltonian in the subspace with conservative excitation number $2$,
whose bases are $\{|0_{q}2_{c}\rangle,|1_{q}1_{c}\rangle,\text{|\ensuremath{2_{q}0_{c}\rangle}\}}$,
is given by 
\begin{equation}
	H_{qc}^{(2)}=\hbar\left[\begin{array}{ccc}
		\Delta(t)-E_{C}^{c}/\hbar & \sqrt{2}g_{qc} & 0\\
		\sqrt{2}g_{qc} & 0 & \sqrt{2}g_{qc}\\
		0 & \sqrt{2}g_{qc} & -\Delta(t)-E_{C}^{q}/\hbar
	\end{array}\right],
\end{equation}
where $E_{c}^{q}$ and $E_{c}^{c}$ are the charging energies of the
qubit and the coupler. Anticrosses of the energy levels of the 
eigenstates take place round $\Delta(t)=E_{C}^{c}/\hbar$ and $\Delta(t)=-E_{C}^{q}/\hbar$.
Similar condition $|\partial_{t}\Delta(t)|=\beta^{\prime}\left((\Delta(t)-\Omega_{C})^{2}+8g_{qc}^{2}\right)^{\frac{3}{2}}/(\sqrt{2}g_{qc})$,
$\beta^{\prime}\ll1$ should preserve near the anticrosses to ensure
adiabaticity, where $\Omega_{C}$ can be $E_{C}^{c}/\hbar$ or $-E_{C}^{q}/\hbar$.
In our experimental setting, $E_{C}^{q}/(2\pi\hbar)=187.4$ MHz and
$E_{C}^{c}/(2\pi\hbar)=330$ MHz. The approximation of the energy
gap at the neighbourhood is valid if the two anticross frequencies
are far apart enough. For simplicity, the frequency trajectory of
the coupler can be derived by only considering the anticross of the
eigenstates $\widetilde{|0_{q}1_{c}\rangle}$ and $\widetilde{|1_{q}0_{c}\rangle}$
at $\Delta(t)=0$, and the adiabaticity can be well preserved as long
as the trajectory varies adequately slow around $\Delta(t)=E_{C}^{c}/\hbar$
and $\Delta(t)=-E_{C}^{q}/\hbar$. Numerical diagonalization of the qubit-coupler Hamiltonian yields the energy spectrum in Fig.~\ref{fig:qcanticross}.

If $\Delta(t)$ satisfies the boundary condition $\Delta(0)=\omega_{0}^\prime=2\pi f_{0}$
and $\Delta(\tau)=\omega_{\tau}=2\pi f_{\tau}$, the solution of Eq.~\ref{eq:adiabatic_condition_01anticross} is 
\begin{equation}
	\Delta(t)=-\frac{8g(\beta gt+\delta)}{\sqrt{1-16(\beta gt)^{2}-32\beta\delta gt-16\delta^{2}}},
\end{equation}
with 
\begin{equation}
	\beta=\frac{-4\delta(\omega_{\tau}^{2}+4g^{2})-\omega_{\tau}\sqrt{\omega_{\tau}^{2}+4g^{2}}}{4g\tau(\omega_{\tau}^{2}+4g^{2})},
\end{equation}
\begin{equation}
	\delta=-\frac{\omega_{0}^\prime}{\sqrt{16\omega_{0}^{\prime 2}+64g^{2}}},
\end{equation}
and $g:=g_{qc}$. For state $|2\rangle$ reset, the solutions can
be obtained by taking the substitutions $g\mapsto\sqrt{2}g_{qc}$
and $\Delta(t)\mapsto\Delta(t)-\Omega_{C}$. The Z pulse amplitude
$Z_{c}(t)$ for the coupler flux control line needed is related to
the coupler frequency $\omega_{c}(t)$ by

\begin{equation}
	Z_{c}^{\pm}(t)=\frac{1}{k}\left[\pm\arcsin\sqrt{\frac{\frac{(\hbar\omega_{c}(t)+E_{C}^{c})^{4}}{(8E_{J}^{c}E_{C}^{c})^{2}}-1}{d^{2}-1}}-b\right],
\end{equation}
where $E_{J}$ is the effective Josephson energy of the SQUID, $d$
is the junction asymmetry and $k$ and $b$ are the effective coefficients
that the magnetic flux $\Phi$ through the SQUID ring is related to
$Z_{c}$ by $\pi\Phi/\Phi_{0}=kZ_{c}+b$.  This waveform enables
faster frequency variation of the coupler when it is far detuned from
the qubit, while slowing down the change rate during proximity. Compared
to ramp functions, this achieves adiabatic exchange in a shorter duration.
As mentioned in the literature, experimentally we treat $g$ as a
tunable parameter to determine. And in conjunction with the coupler's
start and end frequencies $f_{0}$ and $f_{\tau}$ , pulse duration
$\tau$ and the qubit frequency, it determines the frequency trajectory.
Typically, $f_{0}$ and $f_{\tau}$ are predetermined based on operational
requirements, while the adiabaticity of the pulse can be quantified
using the factor $\beta$. As for the qubit frequency in the definition
of $\Delta(t)$, we change it into a tunable center frequency $\omega_{{\rm center}}=2\pi f_{{\rm center}}$,
and the coupler frequency is given by $\omega_{c}(t)=\omega_{{\rm center}}+\Delta(t)$. 

\begin{figure*}[htbp]
	
	\includegraphics[scale=0.55]{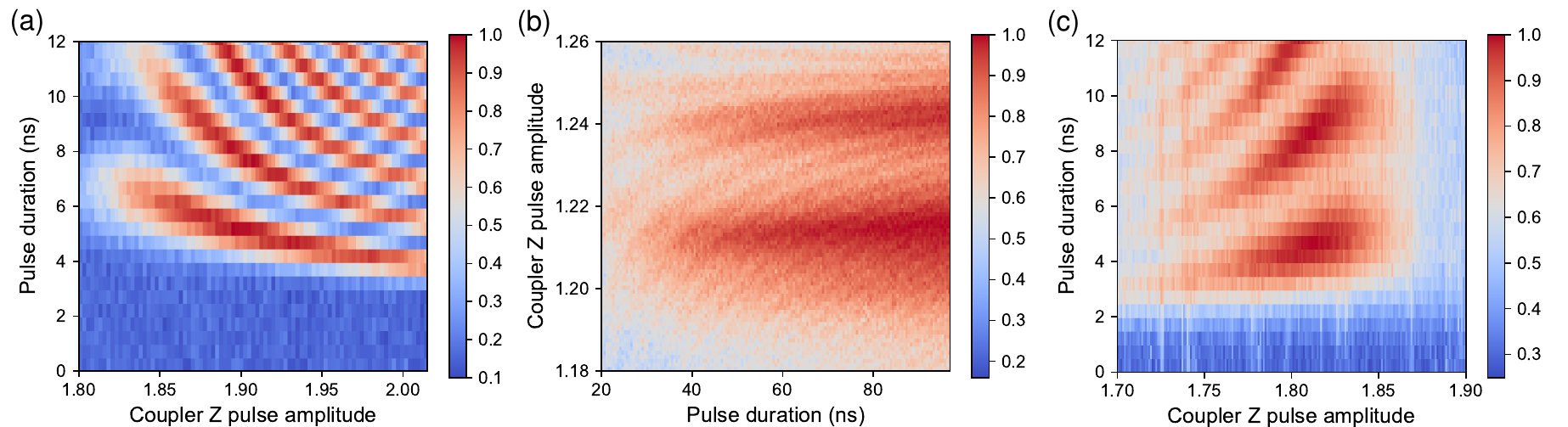}
	
	\caption{Parameter scan results of qubit reset fidelity for resetting state $|1\rangle$ with flat
		square pulse. (a-b) Qubit-coupler swap and coupler-filter swap with square pulse of different coupler Z pulse amplitude and duration. (c) Qubit-coupler swap with the coupler operation point being set to the lower sweet spot.}\label{fig:scan-reset-state-1}
\end{figure*}

The waveform and the reset probability are shown in Fig.~\ref{fig:The-waveforms-and},
~\ref{fig:scan-reset-state-1} and ~\ref{fig:scan-reset-state-21}.
For resonance qubit-coupler exchange with the qubit prepared at state
$|1\rangle$, we choose the qubit-coupler swap duration as $4.5$ ns
and coupler Z pulse amplitude $1.916$ from Fig.~\ref{fig:scan-reset-state-1}(a),
and then perform coupler-filter swap {[}Fig.~\ref{fig:scan-reset-state-1}(b){]}.
We verify the reset of the coupler by swaping the residual excitation
back to the qubit using the same swap duration and Z pulse amplitude.
Owing to the high leakage rate of the Purcell filter, the residual
probability decreases steadily with time when the qubit and the filter
is on resonance, which is over-damped behavior. Selecting swap duration
of $70$ ns and Z pulse amplitude $1.2145$ is sufficient for suppressing
the residual excitation on coupler below $0.5\%$, making it ready
for being reused.

Note that as the reset error approaches zero, the fluctation of probability
from measurement outcomes begin to dominate, necessitating increased
repetition of measurements to achieve more stable results. Apart from
this, we have no direct access to the excitation probability in the
Purcell filter, while the reset error would accumulate if photons
do not actually leak out of the filter before the next reset. To improve
the accuracy and reliability of the characterization of reset error,
we repeat the prepare-state-and-reset pulse sequence multiple times
to enable error accumulation. And we perform scan of reset parameters
with respect to the accumulation of error in average of 3000 shots
to further optimize the reset fidelity. For characterizing the accumulation
of error with the number of repetition, we take the average of 30000
shots for each data point, the result is presented in Fig.~\ref{fig:multiple_prepare_reset_experiment}(a).
The error increases and stabilized around $1.2\%$ for state $|1\rangle$ reset. The mechanism
of the error accumulation is explained in Appendix.~\ref{appH}. 

\begin{figure}[!htb]
	
	\includegraphics[scale=0.6]{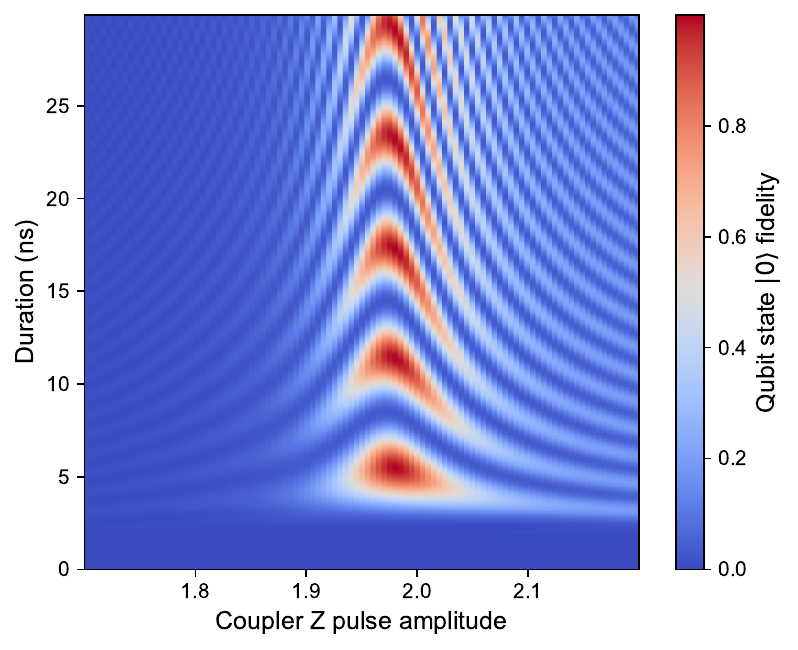}
	
	\caption{Simulated reset fidelity of state $|1\rangle$ with Gaussian filtered
		square pulse.}\label{fig:sim-reset-Gaussian-filt}
\end{figure}
If the coupler is left on its sweet spot during state preparation and readout, we find Landau-Zener-Stuckelberg (LZS) interference fringes~\cite{Kim2025fastunconditional} in QC swap images {[}Fig.~\ref{fig:scan-reset-state-1}(a){]} that even when the flat stage
of the Z pulse pushes the coupler to the lower sweet spot which is
far off resonance with the qubit, the state probability still varies
with time. As is shown in Fig.~\ref{fig:scan-reset-state-1}(c), when
the idling point of the coupler is set to the lower sweet spot, the
fringes of swap far below qubit frequency no longer exist, while it
turns out that fringes occur far above qubit frequency, which is led
by LZS interference. We can reproduce these fringes numerically by
convolving analytical control pulses with a Gaussian kernel of ${\rm FWHM}=1.56~{\rm ns}$,
corresponding to a Gaussian filter function with 3 dB-attenuation bandwith
of approximately $400$ MHz, to make the pulse more slow-varying, as
is shown in Fig.~\ref{fig:sim-reset-Gaussian-filt}, which further
demonstrates that the exchange occurs when the coupler frequency crosses
the qubit frequency. The features in Fig.~\ref{fig:scan-reset-state-1}(a)
also partially originate from waveform distortions other than bandwidth
limitations, typically exhibiting exponential-decay characteristics.
Prior to acquiring data for qubit reset, we perform pulse shape calibration of the coupler's Z channel drive~\cite{li2025high}.

Though we can still perform on-resonance QC swap with high fidelity for state $|1\rangle$
reset, the LZS interference effect undermines the fidelity of the
adiabatic QC swap for resetting both the $|1\rangle$ and $|2\rangle$
population. If we set the coupler's operating point above the qubit
frequency, and start the reset operation by first biasing the coupler
frequency below the qubit frequency, then a portion of the qubit excitations
would be non-adiabatically exchanged to the coupler. This results
in non-negligible occupation in the eigenstates overlapping
predominantly with $|0_{q}1_{c}\rangle$, $|1_{q}1_{c}\rangle$ and
$|0_{q}2_{c}\rangle$ before performing the qubit-coupler swap. After
adiabatic exchange, these eigenstates may primarily overlap with $|1_{q}0_{c}\rangle$, $|1_{q}1_{c}\rangle$ and
$|2_{q}0_{c}\rangle$, respectively [Fig.~\ref{fig:qcanticross}]. In other words, the excitations
would be transitioned back to the qubit. Therefore, we position the
coupler's operating point below the qubit frequency, so that during
each reset, the coupler frequency crosses the qubit frequency only
once. After the coupler-filter swap, since almost no residual excitation
is left in the qubit or the coupler, biasing the coupler back to its
operating point, even when it crosses the qubit frequency,  does not
introduce reset error. 

\begin{figure}[!htb]
	
	\includegraphics[scale=0.62]{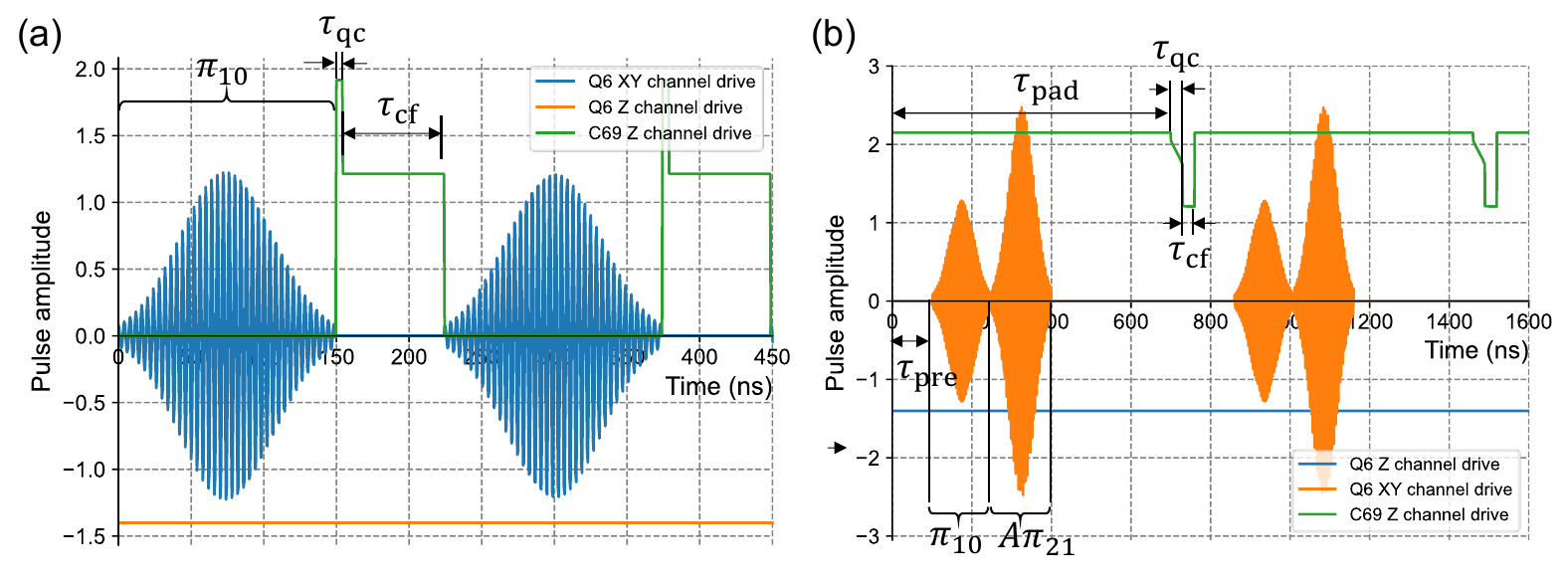}
	
	\caption{Waveforms for repeated state-preparation-and-reset cycles, with definitions of relevant time intervals. (a) In each cycle, the qubit is initialized using a $\pi_{10}$ pulse in the XY channel, followed by reset via sequential qubit-coupler (QC) and coupler-filter (CF) swaps, each implemented with square pulses of duration $\tau_{\rm qc}$ and $\tau_{\rm cf}$ respectively. (b) In each cycle, the qubit state is prepared with $\pi_{10}$ pulse and subsequent $\pi_{21}$ pulse with $A$ being the amplitude scaling factor, and reset with qubit-coupler swap for adiabatic pulse duration $\tau_{\rm qc}$ and then coupler-filter swap for ramp pulse duration $\tau_{\rm cf}$.}\label{fig:The-waveforms-and}
\end{figure}

\begin{figure}[!htb]
	\includegraphics[scale=0.7]{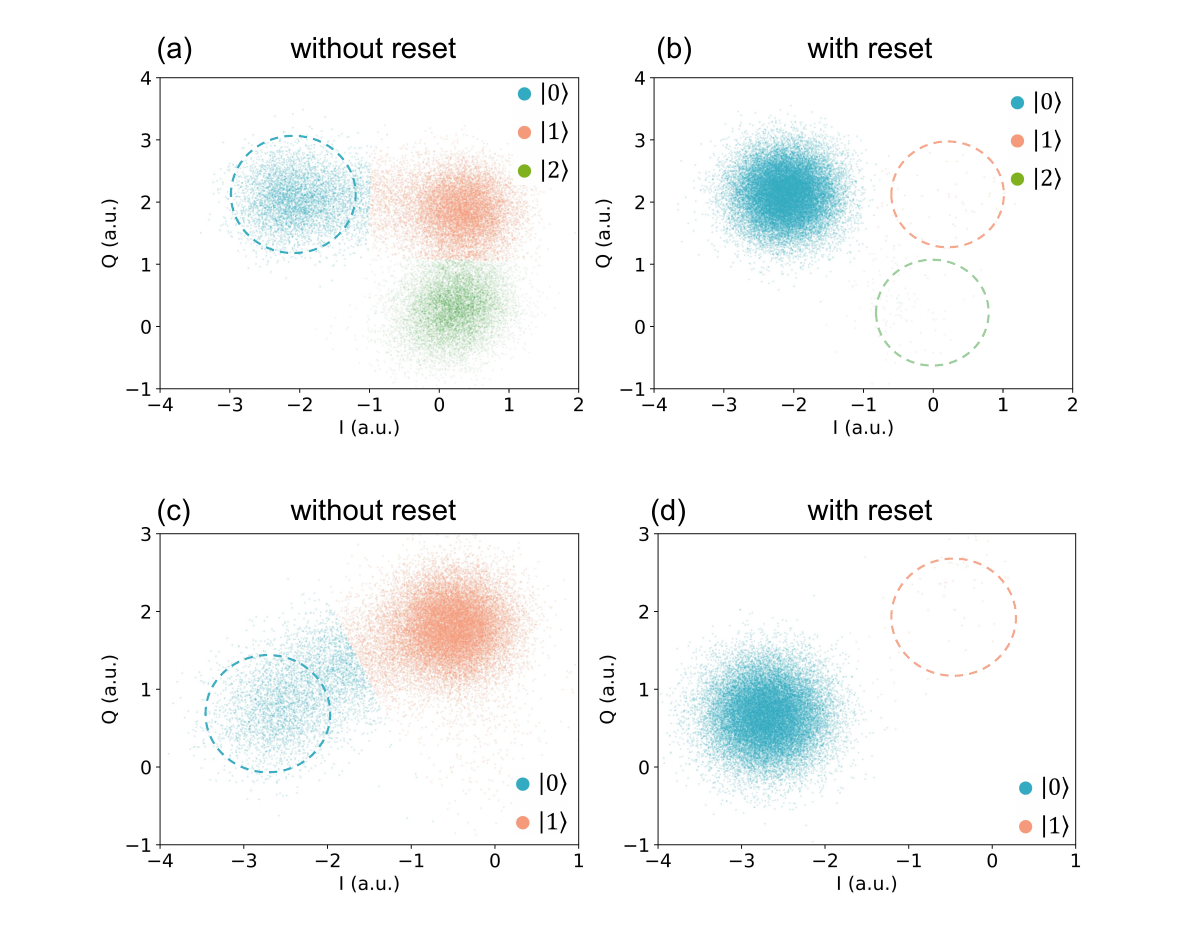}
	\caption{Raw IQ  data with (without) one cycle of reset. (a,b) Initial state with prepared both $\ket{2}$ and $\ket{1}$ excitation. ‌(c,d) Initial state with only $\ket{1}$ excitation.}\label{fig:IQ-raw-reset21}
\end{figure}

\begin{figure*}[htbp]
	
	\includegraphics[scale=0.57]{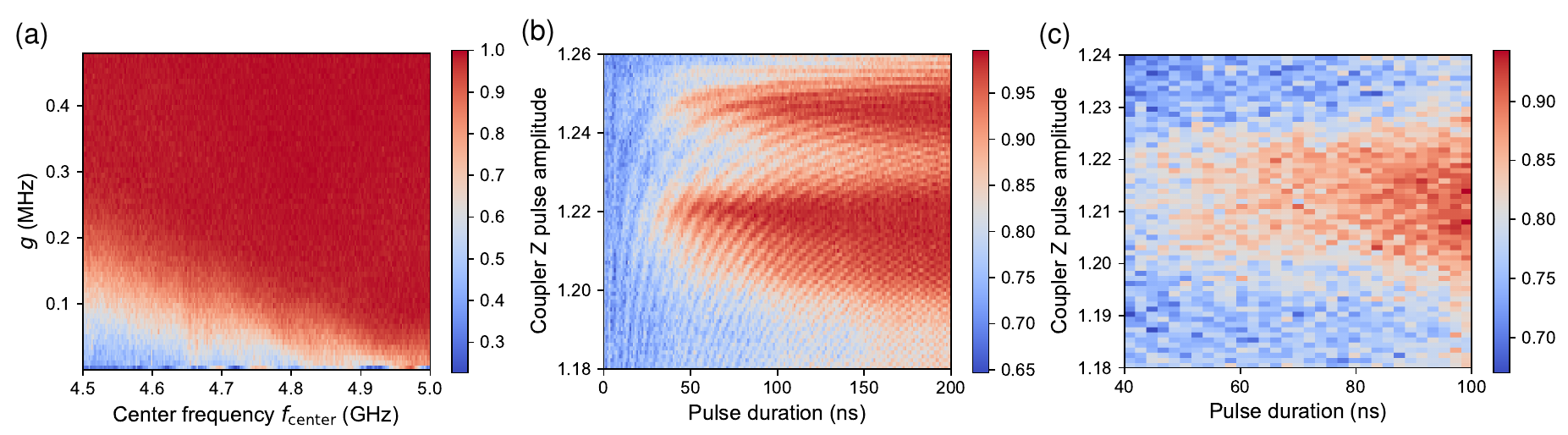}
	
	\caption{Parameter scan of $Q_6$ state $|0\rangle$ probability for resetting both $|1\rangle$ and
		$|2\rangle$ state with adiabatic pulse. (a) Adiabatic qubit-coupler swap with different $g$ and center frequency $f_{\rm center}$. (b) Coupler-filter swap with state preparation and reset process repeated for 20 times. Each state preparation is performed by applying a $\pi_{10}$ pulse. (c) Coupler-filter swap with state preparation and reset process repeated for 20 times, while each state preparation is performed by applying a $\pi_{10}$ pulse and subsequently a $\pi_{21}/2$ pulse.}\label{fig:scan-reset-state-21}
	
\end{figure*}

To avoid affecting the qubit frequency, we set the operating point
of the coupler at the lower sweet spot, with Z amplitude $2.15$
and frequency around $4.0$ GHz. The experimental sequence maintains
a $\tau_{{\rm pad}}=700~{\rm ns}$ interval between successive reset
operations. During this interval, the coupler frequency is stabilized
at the operational point while state preparation is performed using
$\pi_{10}$ and then $\pi_{21}$ pulses with identical $150$ ns durations.
The state preparation commences $\tau_{{\rm pre}}=100$ ns after the
completion of each reset cycle. The waveforms and time intervals are
depicted in Fig.~\ref{fig:The-waveforms-and}. We set the start frequency
to be $4$ GHz and the end frequency $5.5$ GHz. The pulse duration
is set to be $30$ ns. We vary the coefficient $g$ and the center
frequency $f_{{\rm center}}$ to search for optimal qubit-coupler
swap parameter. For finding the coupler-filter swap parameters, we
simply implement prepare-state-and-reset sequence repeated for $20$
times and measure the final qubit state $|0\rangle$ probability.
With coupler-filter swap parameters found, we iteratively optimize
the qubit-coupler and coupler-filter swap parameters using repeated
prepare-state-and-reset sequence. The reset fidelity scan results
are shown in Fig.~\ref{fig:scan-reset-state-21}. To validate that
the protocol resets both $|1\rangle$ and $|2\rangle$ well, we vary
the pulse amplitude for preparing different superpositions of $|1\rangle$
and $|2\rangle$. The results show that for the optimal reset parameters, as the proportion of the
state $|2\rangle$ increases, the reset error rate exhibits only a
minor increase, as is presented in the main text. While it can be seen in Fig.~\ref{fig:scan-reset-state-21}(b-c)
that under identical coupler-filter exchange parameters, the reset
fidelity exhibits discernible discrepancies when the initial states
are prepared in $|1\rangle$ versus $\left(|1\rangle+|2\rangle\right)/\sqrt{2}$.
A longer swap duration is required to achieve low residual error rate
when it involves state $|2\rangle$ occupation. We find optimal parameters
$f_{{\rm center}}=4.86~{\rm GHz}$, $g/(2\pi)=0.605~{\rm MHz}$, $\tau_{\rm qc}=30~{\rm ns}$,
$\tau_{\rm cf}=170~{\rm ns}$, $Z_{\rm c}^{\rm (cf)}=1.209$ and the slope of the coupler-filter swap ramp pulse ${m}_{\rm cf}=-0.00016~/{\rm ns}$. The scatter plot of raw IQ data points used for calculating the reset fidelity is presented in Fig.~\ref{fig:IQ-raw-reset21}.

\begin{figure}[!htb]
	\includegraphics[scale=0.6]{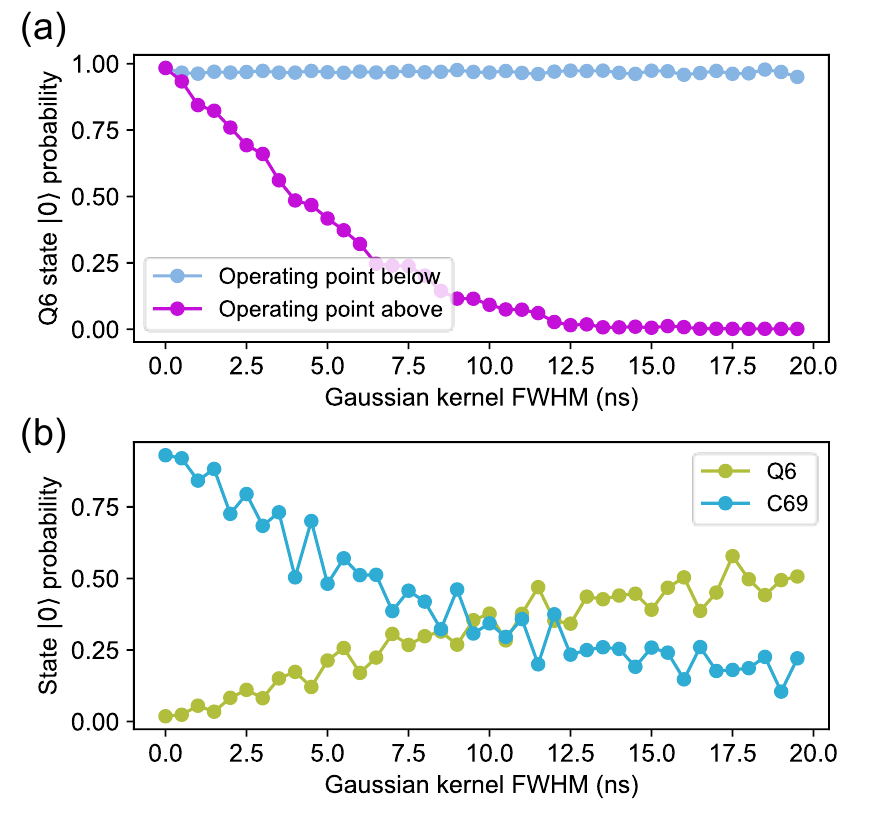}
	\caption{The effect of LZS interference on reset fidelity. (a) Comparison of reset fidelity with the coupler's operating point below and above qubit frequency. (b) The excitation of $Q_6$ transfers to $C_{69}$ during the process of biasing the coupler's frequency below qubit frequency from above, contributing to more reset error as the Gaussian kernel width increases. }\label{fig:LZS-Gaussian-conv}
	
\end{figure}

We numerically illustrate how LZS interference affect the reset fidelity of the adiabatic swap
by convolving the Z pulse with Gaussian kernels of different kernel width {[}Fig.~\ref{fig:LZS-Gaussian-conv}{]}. Using the same pulse parameters while setting the coupler's
operating point below and above qubit frequency respectively, it's
found that in the former case, the reset error rate remains unaffected
by the filter function. This reset protocol demonstrates remarkable
robustness over waveform distortion and bandwith limitations. In the
latter case, however, as the waveform evolves more gradually, the
reset error rate exhibits a substantial increase. By simulating the coupler biasing from its upper sweet spot to lower sweet spot, while staying at the lower sweet spot subsequently, it's found that the excitation transfers from the qubit to the coupler during the process. 

The high leakage rate enables rapid reset and reuse of the Purcell
filter, even if it is shared across multiple couplers for performing
reset operation. The coupler-filter swap is over-damped, and higher
reset fidelity can be reached by allowing slightly longer swap duration [Fig.~\ref{fig:cf_swap_time_tradeoff}],
which is in contrast to the under-damped behavior that can occur when
using qubit-resonator swap, where the coupler excitation can not be
transferred completely at the minimum points of each swap circle due
to the dissipation and we have discrete time points to choose from [Fig.~\ref{fig:multiple_prepare_reset_experiment}(b)].
With this, the coupler-filter swap is theoretically
more efficient since the leakage rate of the resonator is small,
which should be able to match $2\chi$ while $\chi$ is designed to be
small to protect the qubit from decoherence, and using coupler-resonator
swap would potentially introduce larger qubit dephasing rate over
a significant duration.

\begin{figure}[!htb]
	\includegraphics[scale=0.6]{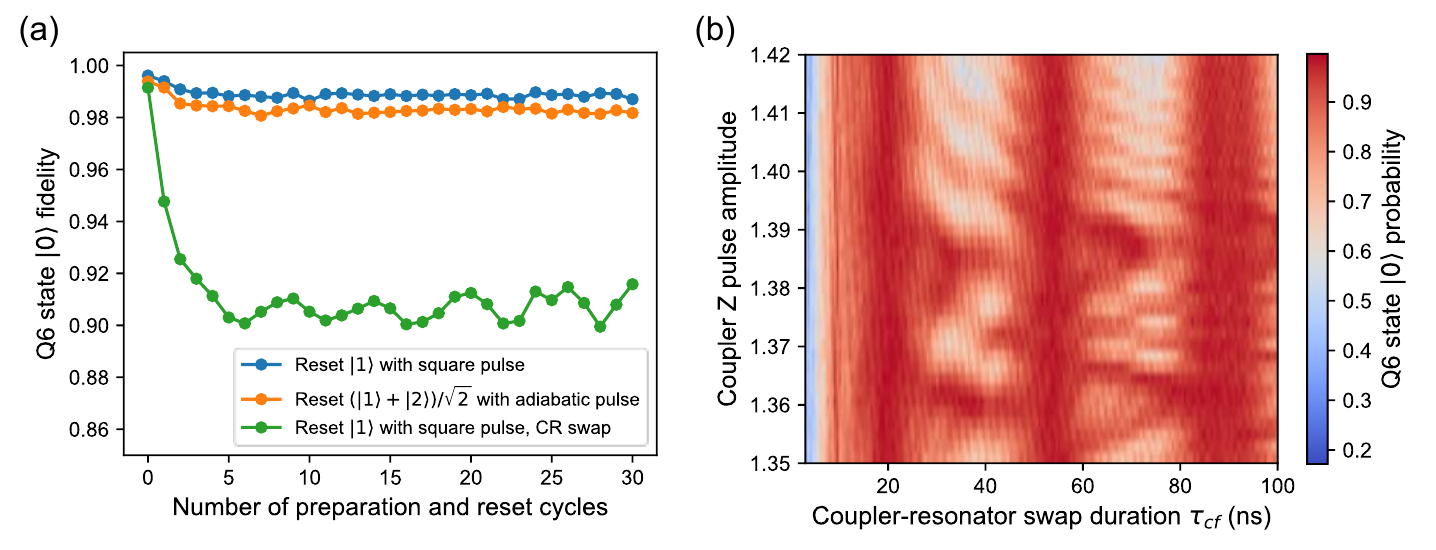}
	\caption{(a) Reset error accumulation with the number of state preparation and reset cycles. The qubit states are reset with square pulse and CF swap, adiabatic pulse and CF swap, and square pulse but CR (coupler-resonator) swap, where the readout filter Z pulse amplitude is set to $1.36$ during the swap. (b) Coupler reset fidelity with respect to qubit with 1 reset cycle. The coupler is reset with the resonator, and the residual excitation is measured by swapping back the excitation to the qubit. The filter Z pulse amplitude is set to 1.5 to reach high resonator leakage rate.} \label{fig:multiple_prepare_reset_experiment}
\end{figure}

\begin{figure}[!htb]
	\includegraphics[scale=0.65]{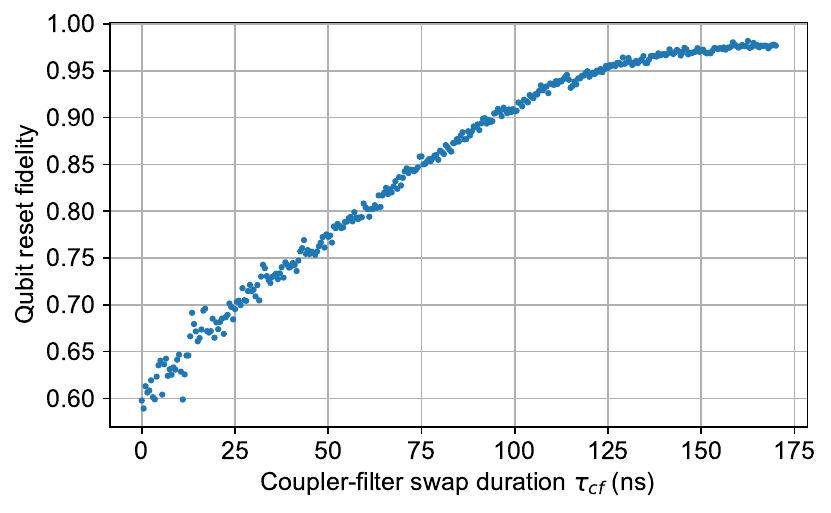}
	\caption{The reset fidelity after 20 state-preparation-and-reset cycles increases with the coupler-filter swap duration monotonically.}\label{fig:cf_swap_time_tradeoff}
\end{figure}

\begin{figure}[!htb]
	\includegraphics[scale=0.15]{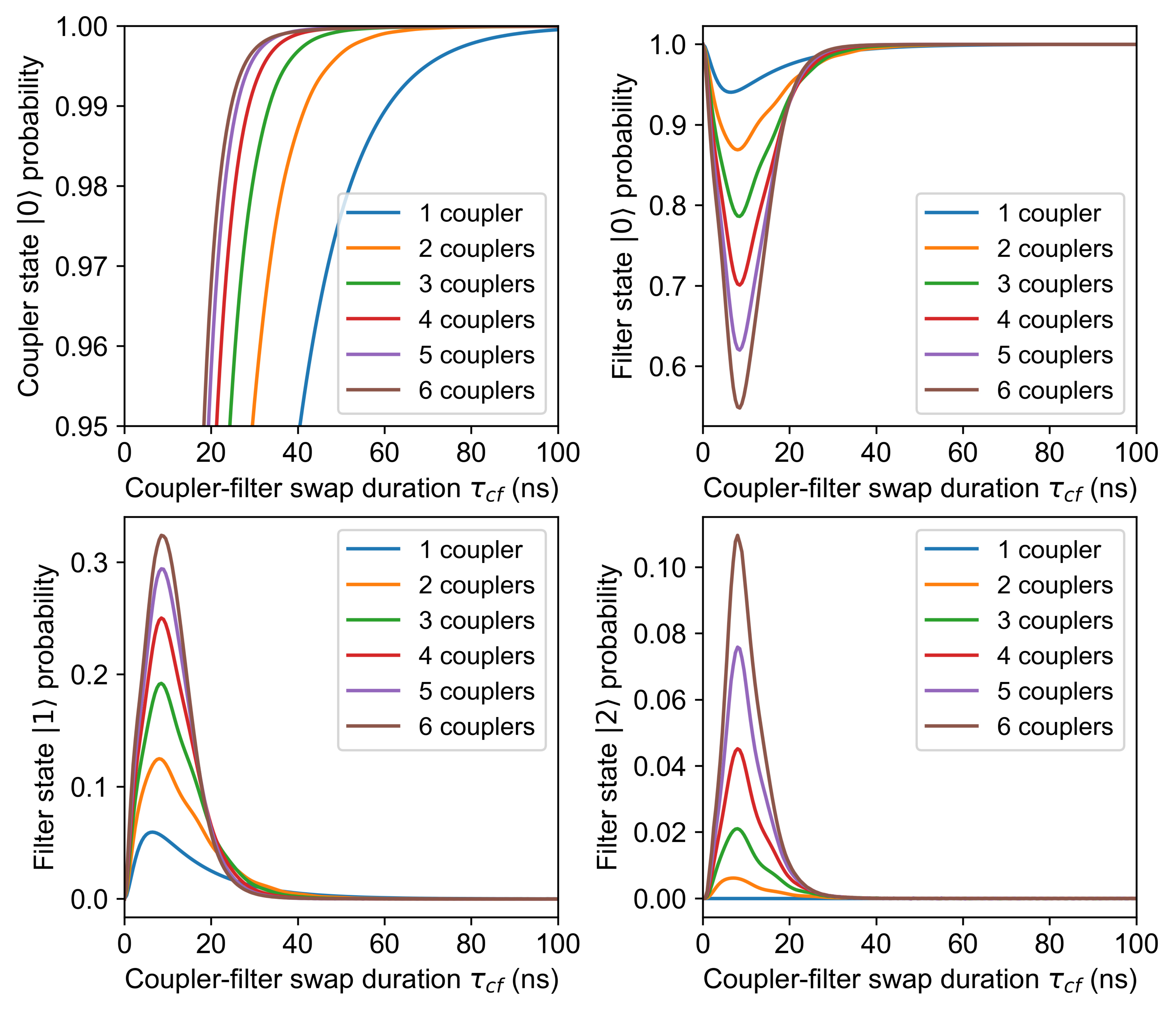}
	\caption{Different numbers of couplers performing coupler-filter swap with a common Purcell filter simultaneously. The couplers are designed to be identical and the initial states of the couplers are all prepared to be $|1\rangle$.}\label{fig:multiple-coupler-reset}
\end{figure}

To validate the scalability of the coupler-filter coupling architecture, we numerically simulate the situation where multiple couplers simultaneously perform resonant exchange with a common Purcell filter of frequency $\omega_f/(2 \pi)= 7.01367~\rm{GHz}$ and $\kappa_f=140~\rm{MHz}$, the result of which is shown in Fig. \ref{fig:multiple-coupler-reset}. The couplers are assumed to be identical, with the coupler-filter coupling strength $g_{cf}/(2\pi)=20~\rm{MHz}$ at this frequency. They are all prepared in state $|1\rangle$ and then set to be resonant with the filter. As the number of number of couplers increases, the reset process of each coupler becomes faster, due to the fact that the qubit excitations are swapped to higher energy levels of the Purcell filter with low anharmonicity, where the corresponding state decay rate is larger. Therefore, as the excitations accumulate to higer levels, each excitation obtain a larger average decay rate.

\begin{figure}[!htb]
	\includegraphics[scale=0.55]{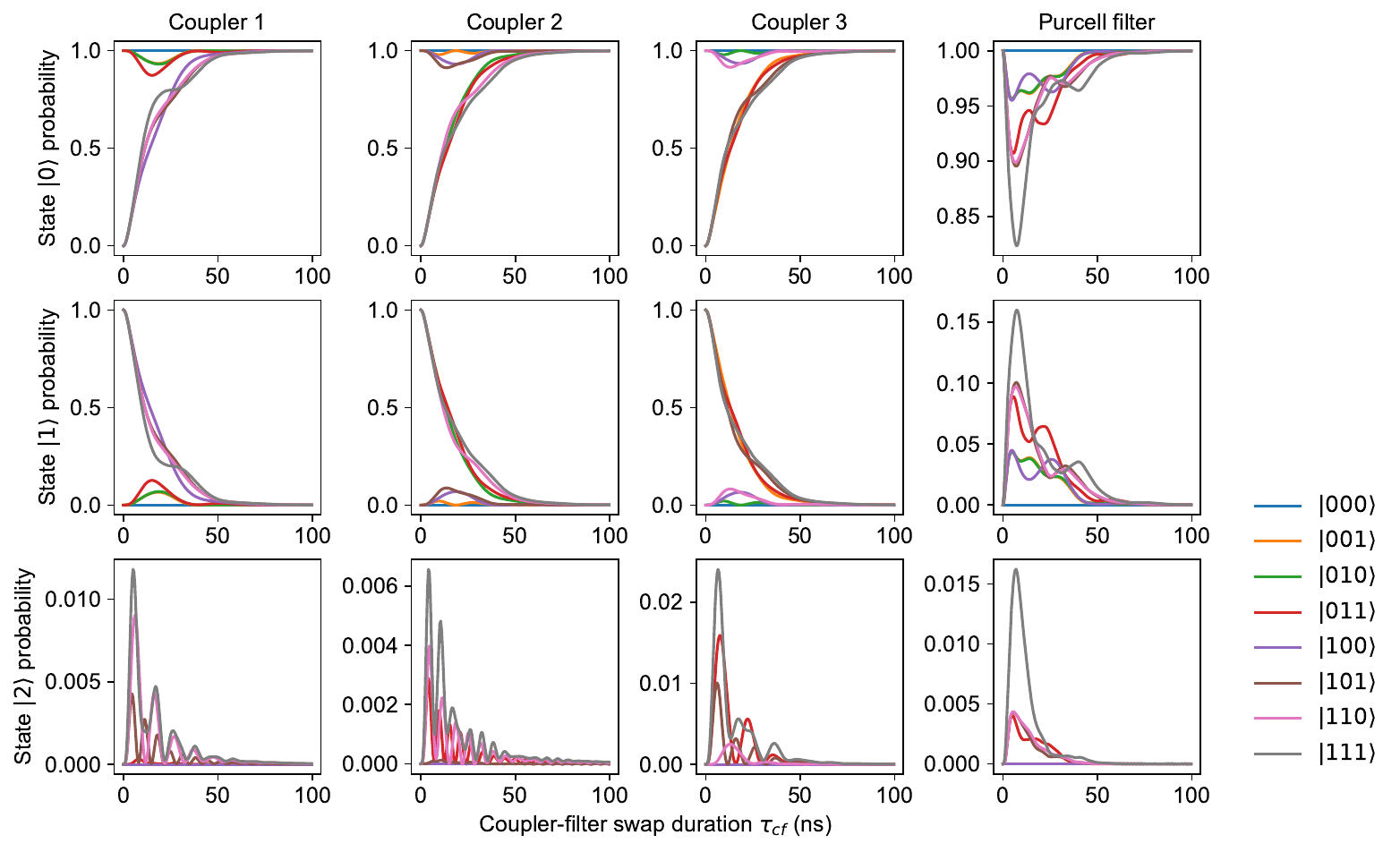}
	\caption{Three couplers simultaneously engage a shared Purcell filter, with different three-coupler initial states prepared.}\label{fig:partially-excited-3c-f-swap}
\end{figure}

A more common situation is that only part of the qubits are excited or the qubit states are in superpositions of $|0\rangle$ and the excited states. When the couplers are partially excited, the on-resonance exchange will reach a stage where the excitations will persist in the couplers without dissipation, which can be explained that the vectorized Liouville superoperator of the Lindbladian form ${\rm vec}(\mathcal{L})= -i (I\otimes H)/\hbar + i(H^T\otimes I)/\hbar+ \kappa(a_f^* \otimes a_f - \frac{1}{2} I \otimes a_f^\dagger a_f - \frac{1}{2} a_f^T a_f^* \otimes I)$ of the system has eigenstates whose corresponding eigenvalues contain approximately zero real part and there is no excited state population of the filter. When the initial state is prepared to overlap with them, these parts do not dissipate. It poses a challenge for simultaneously resetting all of the couplers. However, the non-dissipative eigenstates can be removed by detuning the couplers slightly. As is shown numerically in Fig.~\ref{fig:partially-excited-3c-f-swap}, three couplers are detuned from the Purcell filter by $\Delta_1=0.00~\rm{MHz}$, $\Delta_2=-25.89~\rm{MHz}$ and $\Delta_3=26.74~\rm{MHz}$, enabling fast reset for all initial state configurations. Future research may explore frequency modulation techniques to prevent the emergence of non-dissipative eigenstates and optimize for faster dissipation.

\section{\label{appH}Analysis of reset error accumulation}

We posit that the error accumulation during multiple reset operations
primarily evaluates the effectiveness of the decay of the Purcell
filter. This is because if the coupler-qubit exchange achieves unconditional
reset, even imperfect qubit reset would at most introduce errors in
subsequent state preparation, potentially resulting in lower population
of $|2\rangle$ and $|1\rangle$ states compared to scenarios without
reset errors. Conversely, in the coupler-qubit exchange process, incomplete
coupler reset would lead to residual excitations being transferred
back to the qubit. With the residual excitation accumulates in the
Purcell filter, the residual excitation of the coupler after the coupler-filter
swap increases, which finally induces accumulation in qubit reset
error. 

\begin{figure*}[!htb]
	\includegraphics[scale=0.5]{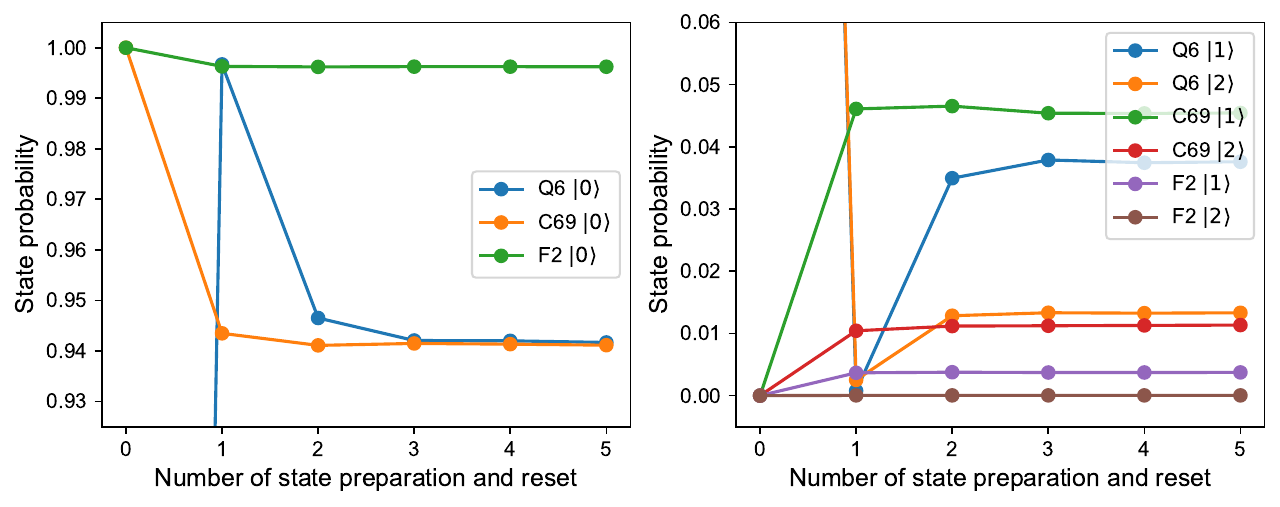}
	\caption{Numerical demonstration of reset error accumulation due to imperfect coupler-filter swap.}\label{fig:adia-multi-prep-reset}
\end{figure*}

We reproduce the phenomena numerically and further illustruate
this mechanism, as shown in Fig.~\ref{fig:adia-multi-prep-reset}. The coupler-filter swap duration is intentially reduce to $70~\rm{ns}$ to introduce larger reset error of the coupler, and the filter dissipation during qubit state preparation is omitted. 
After the first preparation and reset operation, even though the qubit is reset with high fidelity with adiabatic qubit-coupler swap, the coupler's residual excitation is significant, which is swapped back to the qubit in the next reset cycle. As the filter does not completely dissipate to thermal equilibrium, the coupler-filter swap in the second reset cycle has larger coupler reset error. The residual excitation of the coupler is progressively swapped to the qubit, thus leading to accumulation of the qubit's reset error. While the state preparation introduces more excitations, the amount of excitation that coupler exchanges to the filter increase due to larger excited state occupation and therefore larger exchange rate. This two mechanism competes and finally lead to equilibrium, resulting in stable reset error in the following reset cycles.

The accumulation of reset error can be reduced if we perform multiple
reset after each state preparation, as is presented in the main text and can also be reduced predominantly by performing longer coupler-resonator swap [Fig.~\ref{fig:cf_swap_time_tradeoff}]. We can
make a trade off between reset time and reset fidelity, depending
on experimental requirements.  

\newpage
\bibliography{ref}  